%
%
%
%

\ifx\mnmacrosloaded\undefined 
%
%
%
%

\catcode `\@=11 

\def\@version{1.6}
\def\@verdate{18th September 1995}

%
%


\newif\ifprod@font

\ifx\@typeface\undefined
  \def\@typeface{Comp. Modern}\prod@fontfalse
\else
  \prod@fonttrue 
\fi

\def\newfam{\alloc@8\fam\chardef\sixt@@n} 

\ifprod@font
\font\fiverm=mtr10 at 5pt
\font\fivebf=mtbx10 at 5pt
\font\fiveit=mtti10 at 5pt
\font\fivesl=mtsl10 at 5pt
\font\fivett=cmtt8 at 5pt     \hyphenchar\fivett=-1
\font\fivecsc=mtcsc10 at 5pt
\font\fivesf=mtss10 at 5pt
\font\fivei=mtmi10 at 5pt      \skewchar\fivei='177
\font\fivesy=mtsy10 at 5pt     \skewchar\fivesy='60

\font\sixrm=mtr10 at 6pt
\font\sixbf=mtbx10 at 6pt
\font\sixit=mtti10 at 6pt
\font\sixsl=mtsl10 at 6pt
\font\sixtt=cmtt8 at 6pt      \hyphenchar\sixtt=-1
\font\sixcsc=mtcsc10 at 6pt
\font\sixsf=mtss10 at 6pt
\font\sixi=mtmi10 at 6pt       \skewchar\sixi='177
\font\sixsy=mtsy10 at 6pt      \skewchar\sixsy='60

\font\sevenrm=mtr10 at 7pt
\font\sevenbf=mtbx10 at 7pt
\font\sevenit=mtti10 at 7pt
\font\sevensl=mtsl10 at 7pt
\font\seventt=cmtt8 at 7pt     \hyphenchar\seventt=-1
\font\sevencsc=mtcsc10 at 7pt
\font\sevensf=mtss10 at 7pt
\font\seveni=mtmi10 at 7pt      \skewchar\seveni='177
\font\sevensy=mtsy10 at 7pt     \skewchar\sevensy='60

\font\eightrm=mtr10 at 8pt
\font\eightbf=mtbx10 at 8pt
\font\eightit=mtti10 at 8pt
\font\eighti=mtmi10 at 8pt      \skewchar\eighti='177
\font\eightsy=mtsy10 at 8pt     \skewchar\eightsy='60
\font\eightsl=mtsl10 at 8pt
\font\eighttt=cmtt8             \hyphenchar\eighttt=-1
\font\eightcsc=mtcsc10 at 8pt
\font\eightsf=mtss10 at 8pt

\font\ninerm=mtr10 at 9pt
\font\ninebf=mtbx10 at 9pt
\font\nineit=mtti10 at 9pt
\font\ninei=mtmi10 at 9pt      \skewchar\ninei='177
\font\ninesy=mtsy10 at 9pt     \skewchar\ninesy='60
\font\ninesl=mtsl10 at 9pt
\font\ninett=cmtt9             \hyphenchar\ninett=-1
\font\ninecsc=mtcsc10 at 9pt
\font\ninesf=mtss10 at 9pt

\font\tenrm=mtr10
\font\tenbf=mtbx10
\font\tenit=mtti10
\font\teni=mtmi10		\skewchar\teni='177
\font\tensy=mtsy10		\skewchar\tensy='60
\font\tenex=cmex10
\font\tensl=mtsl10
\font\tentt=cmtt10		\hyphenchar\tentt=-1
\font\tencsc=mtcsc10
\font\tensf=mtss10

\font\elevenrm=mtr10 at 11pt
\font\elevenbf=mtbx10 at 11pt
\font\elevenit=mtti10 at 11pt
\font\eleveni=mtmi10 at 11pt      \skewchar\eleveni='177
\font\elevensy=mtsy10 at 11pt     \skewchar\elevensy='60
\font\elevensl=mtsl10 at 11pt
\font\eleventt=cmtt10 at 11pt     \hyphenchar\eleventt=-1
\font\elevencsc=mtcsc10 at 11pt
\font\elevensf=mtss10 at 11pt

\font\twelverm=mtr10 at 12pt
\font\twelvebf=mtbx10 at 12pt
\font\twelveit=mtti10 at 12pt
\font\twelvesl=mtsl10 at 12pt
\font\twelvett=cmtt12             \hyphenchar\twelvett=-1
\font\twelvecsc=mtcsc10 at 12pt
\font\twelvesf=mtss10 at 12pt
\font\twelvei=mtmi10 at 12pt      \skewchar\twelvei='177
\font\twelvesy=mtsy10 at 12pt     \skewchar\twelvesy='60

\font\fourteenrm=mtr10 at 14pt
\font\fourteenbf=mtbx10 at 14pt
\font\fourteenit=mtti10 at 14pt
\font\fourteeni=mtmi10 at 14pt      \skewchar\fourteeni='177
\font\fourteensy=mtsy10 at 14pt     \skewchar\fourteensy='60
\font\fourteensl=mtsl10 at 14pt
\font\fourteentt=cmtt12 at 14pt     \hyphenchar\fourteentt=-1
\font\fourteencsc=mtcsc10 at 14pt
\font\fourteensf=mtss10 at 14pt

\font\seventeenrm=mtr10 at 17pt
\font\seventeenbf=mtbx10 at 17pt
\font\seventeenit=mtti10 at 17pt
\font\seventeeni=mtmi10 at 17pt      \skewchar\seventeeni='177
\font\seventeensy=mtsy10 at 17pt     \skewchar\seventeensy='60
\font\seventeensl=mtsl10 at 17pt
\font\seventeentt=cmtt12 at 17pt     \hyphenchar\seventeentt=-1
\font\seventeencsc=mtcsc10 at 17pt
\font\seventeensf=mtss10 at 17pt
\else
\font\fiverm=cmr5
\font\fivei=cmmi5             \skewchar\fivei='177
\font\fivesy=cmsy5            \skewchar\fivesy='60
\font\fivebf=cmbx5

\font\sixrm=cmr6
\font\sixi=cmmi6             \skewchar\sixi='177
\font\sixsy=cmsy6            \skewchar\sixsy='60
\font\sixbf=cmbx6

\font\sevenrm=cmr7
\font\sevenit=cmti7
\font\seveni=cmmi7             \skewchar\seveni='177
\font\sevensy=cmsy7            \skewchar\sevensy='60
\font\sevenbf=cmbx7

\font\eightrm=cmr8
\font\eightbf=cmbx8
\font\eightit=cmti8
\font\eighti=cmmi8			\skewchar\eighti='177
\font\eightsy=cmsy8			\skewchar\eightsy='60
\font\eightsl=cmsl8
\font\eighttt=cmtt8			\hyphenchar\eighttt=-1
\font\eightcsc=cmcsc10 at 8pt
\font\eightsf=cmss8

\font\ninerm=cmr9
\font\ninebf=cmbx9
\font\nineit=cmti9
\font\ninei=cmmi9			\skewchar\ninei='177
\font\ninesy=cmsy9			\skewchar\ninesy='60
\font\ninesl=cmsl9
\font\ninett=cmtt9			\hyphenchar\ninett=-1
\font\ninecsc=cmcsc10 at 9pt
\font\ninesf=cmss9

\font\tenrm=cmr10
\font\tenbf=cmbx10
\font\tenit=cmti10
\font\teni=cmmi10		\skewchar\teni='177
\font\tensy=cmsy10		\skewchar\tensy='60
\font\tenex=cmex10
\font\tensl=cmsl10
\font\tentt=cmtt10		\hyphenchar\tentt=-1
\font\tencsc=cmcsc10
\font\tensf=cmss10

\font\elevenrm=cmr10 scaled \magstephalf
\font\elevenbf=cmbx10 scaled \magstephalf
\font\elevenit=cmti10 scaled \magstephalf
\font\eleveni=cmmi10 scaled \magstephalf	\skewchar\eleveni='177
\font\elevensy=cmsy10 scaled \magstephalf	\skewchar\elevensy='60
\font\elevensl=cmsl10 scaled \magstephalf
\font\eleventt=cmtt10 scaled \magstephalf	\hyphenchar\eleventt=-1
\font\elevencsc=cmcsc10 scaled \magstephalf
\font\elevensf=cmss10 scaled \magstephalf

\font\twelverm=cmr10 scaled \magstep1
\font\twelvebf=cmbx10 scaled \magstep1
\font\twelvei=cmmi10 scaled \magstep1      \skewchar\twelvei='177
\font\twelvesy=cmsy10 scaled \magstep1     \skewchar\twelvesy='60

\font\fourteenrm=cmr10 scaled \magstep2
\font\fourteenbf=cmbx10 scaled \magstep2
\font\fourteenit=cmti10 scaled \magstep2
\font\fourteeni=cmmi10 scaled \magstep2		\skewchar\fourteeni='177
\font\fourteensy=cmsy10 scaled \magstep2	\skewchar\fourteensy='60
\font\fourteensl=cmsl10 scaled \magstep2
\font\fourteentt=cmtt10 scaled \magstep2	\hyphenchar\fourteentt=-1
\font\fourteencsc=cmcsc10 scaled \magstep2
\font\fourteensf=cmss10 scaled \magstep2

\font\seventeenrm=cmr10 scaled \magstep3
\font\seventeenbf=cmbx10 scaled \magstep3
\font\seventeenit=cmti10 scaled \magstep3
\font\seventeeni=cmmi10 scaled \magstep3	\skewchar\seventeeni='177
\font\seventeensy=cmsy10 scaled \magstep3	\skewchar\seventeensy='60
\font\seventeensl=cmsl10 scaled \magstep3
\font\seventeentt=cmtt10 scaled \magstep3	\hyphenchar\seventeentt=-1
\font\seventeencsc=cmcsc10 scaled \magstep3
\font\seventeensf=cmss10 scaled \magstep3
\fi

\def\hexnumber#1{\ifcase#1 0\or1\or2\or3\or4\or5\or6\or7\or8\or9\or
  A\or B\or C\or D\or E\or F\fi}

\def\makestrut{%
  \setbox\strutbox=\hbox{%
    \vrule height.7\baselineskip depth.3\baselineskip width \z@}%
}

\def\baselinestretch{1}
\newskip\tmp@bls

\def\b@ls#1{
  \tmp@bls=#1\relax
  \baselineskip=#1\relax\makestrut
  \normalbaselineskip=\baselinestretch\tmp@bls
  \normalbaselines
}

\def\nostb@ls#1{
  \normalbaselineskip=#1\relax
  \normalbaselines
  \makestrut
}

%

\newfam\scfam  
\newfam\sffam  

\def\mit{\fam\@ne}
\def\cal{\fam\tw@}
\def\em{\ifdim\fontdimen1\font>\z@ \rm\else\it\fi}

\textfont3=\tenex
\scriptfont3=\tenex
\scriptscriptfont3=\tenex

\setbox0=\hbox{\tenex B} \p@renwd=\wd0 

\def\eightpoint{
  \def\rm{\fam0\eightrm}%
  \textfont0=\eightrm \scriptfont0=\sixrm \scriptscriptfont0=\fiverm%
  \textfont1=\eighti  \scriptfont1=\sixi  \scriptscriptfont1=\fivei%
  \textfont2=\eightsy \scriptfont2=\sixsy \scriptscriptfont2=\fivesy%
  \textfont\itfam=\eightit\def\it{\fam\itfam\eightit}%
  \ifprod@font
    \scriptfont\itfam=\sixit
      \scriptscriptfont\itfam=\fiveit
  \else
    \scriptfont\itfam=\eightit
      \scriptscriptfont\itfam=\eightit
  \fi
  \textfont\bffam=\eightbf%
    \scriptfont\bffam=\sixbf%
      \scriptscriptfont\bffam=\fivebf%
  \def\bf{\fam\bffam\eightbf}%
  \textfont\slfam=\eightsl\def\sl{\fam\slfam\eightsl}%
  \ifprod@font
    \scriptfont\slfam=\sixsl
      \scriptscriptfont\slfam=\fivesl
  \else
    \scriptfont\slfam=\eightsl
      \scriptscriptfont\slfam=\eightsl
  \fi
  \textfont\ttfam=\eighttt\def\tt{\fam\ttfam\eighttt}%
  \ifprod@font
    \scriptfont\ttfam=\sixtt
      \scriptscriptfont\ttfam=\fivett
  \else
    \scriptfont\ttfam=\eighttt
      \scriptscriptfont\ttfam=\eighttt
  \fi
  \textfont\scfam=\eightcsc\def\sc{\fam\scfam\eightcsc}%
  \ifprod@font
    \scriptfont\scfam=\sixcsc
      \scriptscriptfont\scfam=\fivecsc
  \else
    \scriptfont\scfam=\eightcsc
      \scriptscriptfont\scfam=\eightcsc
  \fi
  \textfont\sffam=\eightsf\def\sf{\fam\sffam\eightsf}%
  \ifprod@font
    \scriptfont\sffam=\sixsf
      \scriptscriptfont\sffam=\fivesf
  \else
    \scriptfont\sffam=\eightsf
      \scriptscriptfont\sffam=\eightsf
  \fi
  \def\oldstyle{\fam\@ne\eighti}%
  \b@ls{10pt}\rm\@viiipt%
}
\def\@viiipt{}

\def\ninepoint{
  \def\rm{\fam0\ninerm}%
  \textfont0=\ninerm \scriptfont0=\sixrm \scriptscriptfont0=\fiverm%
  \textfont1=\ninei  \scriptfont1=\sixi  \scriptscriptfont1=\fivei%
  \textfont2=\ninesy \scriptfont2=\sixsy \scriptscriptfont2=\fivesy%
  \textfont\itfam=\nineit\def\it{\fam\itfam\nineit}%
  \ifprod@font
    \scriptfont\itfam=\sixit
      \scriptscriptfont\itfam=\fiveit
  \else
    \scriptfont\itfam=\nineit
      \scriptscriptfont\itfam=\nineit
  \fi
  \textfont\bffam=\ninebf%
    \scriptfont\bffam=\sixbf%
      \scriptscriptfont\bffam=\fivebf%
  \def\bf{\fam\bffam\ninebf}%
  \textfont\slfam=\ninesl\def\sl{\fam\slfam\ninesl}%
  \ifprod@font
    \scriptfont\slfam=\sixsl
      \scriptscriptfont\slfam=\fivesl
  \else
    \scriptfont\slfam=\ninesl
      \scriptscriptfont\slfam=\ninesl
  \fi
  \textfont\ttfam=\ninett\def\tt{\fam\ttfam\ninett}%
  \ifprod@font
    \scriptfont\ttfam=\sixtt
      \scriptscriptfont\ttfam=\fivett
  \else
    \scriptfont\ttfam=\ninett
      \scriptscriptfont\ttfam=\ninett
  \fi
  \textfont\scfam=\ninecsc\def\sc{\fam\scfam\ninecsc}%
  \ifprod@font
    \scriptfont\scfam=\sixcsc
      \scriptscriptfont\scfam=\fivecsc
  \else
    \scriptfont\scfam=\ninecsc
      \scriptscriptfont\scfam=\ninecsc
  \fi
  \textfont\sffam=\ninesf\def\sf{\fam\sffam\ninesf}%
  \ifprod@font
    \scriptfont\sffam=\sixsf
      \scriptscriptfont\sffam=\fivesf
  \else
    \scriptfont\sffam=\ninesf
      \scriptscriptfont\sffam=\ninesf
  \fi
  \def\oldstyle{\fam\@ne\ninei}%
  \b@ls{\TextLeading plus \Feathering}\rm\@ixpt%
}
\def\@ixpt{}

\def\tenpoint{
  \def\rm{\fam0\tenrm}%
  \textfont0=\tenrm \scriptfont0=\sevenrm \scriptscriptfont0=\fiverm%
  \textfont1=\teni  \scriptfont1=\seveni  \scriptscriptfont1=\fivei%
  \textfont2=\tensy \scriptfont2=\sevensy \scriptscriptfont2=\fivesy%
  \textfont\itfam=\tenit\def\it{\fam\itfam\tenit}%
  \ifprod@font
    \scriptfont\itfam=\sevenit
      \scriptscriptfont\itfam=\fiveit
  \else
    \scriptfont\itfam=\tenit
      \scriptscriptfont\itfam=\tenit
  \fi
  \textfont\bffam=\tenbf%
    \scriptfont\bffam=\sevenbf%
      \scriptscriptfont\bffam=\fivebf%
  \def\bf{\fam\bffam\tenbf}%
  \textfont\slfam=\tensl\def\sl{\fam\slfam\tensl}%
  \ifprod@font
    \scriptfont\slfam=\sevensl
      \scriptscriptfont\slfam=\fivesl
  \else
    \scriptfont\slfam=\tensl
      \scriptscriptfont\slfam=\tensl
  \fi
  \textfont\ttfam=\tentt\def\tt{\fam\ttfam\tentt}%
  \ifprod@font
    \scriptfont\ttfam=\seventt
      \scriptscriptfont\ttfam=\fivett
  \else
    \scriptfont\ttfam=\tentt
      \scriptscriptfont\ttfam=\tentt
  \fi
  \textfont\scfam=\tencsc\def\sc{\fam\scfam\tencsc}%
  \ifprod@font
    \scriptfont\scfam=\sevencsc
      \scriptscriptfont\scfam=\fivecsc
  \else
    \scriptfont\scfam=\tencsc
      \scriptscriptfont\scfam=\tencsc
  \fi
  \textfont\sffam=\tensf\def\sf{\fam\sffam\tensf}%
  \ifprod@font
    \scriptfont\sffam=\sevensf
      \scriptscriptfont\sffam=\fivesf
  \else
    \scriptfont\sffam=\tensf
      \scriptscriptfont\sffam=\tensf
  \fi
  \def\oldstyle{\fam\@ne\teni}%
  \b@ls{11pt}\rm\@xpt%
}
\def\@xpt{}

\def\elevenpoint{
  \def\rm{\fam0\elevenrm}%
  \textfont0=\elevenrm \scriptfont0=\eightrm \scriptscriptfont0=\sixrm%
  \textfont1=\eleveni  \scriptfont1=\eighti  \scriptscriptfont1=\sixi%
  \textfont2=\elevensy \scriptfont2=\eightsy \scriptscriptfont2=\sixsy%
  \textfont\itfam=\elevenit\def\it{\fam\itfam\elevenit}%
  \ifprod@font
    \scriptfont\itfam=\eightit
      \scriptscriptfont\itfam=\sixit
  \else
    \scriptfont\itfam=\elevenit
      \scriptscriptfont\itfam=\elevenit
  \fi
  \textfont\bffam=\elevenbf%
    \scriptfont\bffam=\eightbf%
      \scriptscriptfont\bffam=\sixbf%
  \def\bf{\fam\bffam\elevenbf}%
  \textfont\slfam=\elevensl\def\sl{\fam\slfam\elevensl}%
  \ifprod@font
    \scriptfont\slfam=\eightsl
      \scriptscriptfont\slfam=\sixsl
  \else
    \scriptfont\slfam=\elevensl
      \scriptscriptfont\slfam=\elevensl
  \fi
  \textfont\ttfam=\eleventt\def\tt{\fam\ttfam\eleventt}%
  \ifprod@font
    \scriptfont\ttfam=\eighttt
      \scriptscriptfont\ttfam=\sixtt
  \else
    \scriptfont\ttfam=\eleventt
      \scriptscriptfont\ttfam=\eleventt
  \fi
  \textfont\scfam=\elevencsc\def\sc{\fam\scfam\elevencsc}%
  \ifprod@font
    \scriptfont\scfam=\eightcsc
      \scriptscriptfont\scfam=\sixcsc
  \else
    \scriptfont\scfam=\elevencsc
      \scriptscriptfont\scfam=\elevencsc
  \fi
  \textfont\sffam=\elevensf\def\sf{\fam\sffam\elevensf}%
  \ifprod@font
    \scriptfont\sffam=\eightsf
      \scriptscriptfont\sffam=\sixsf
  \else
    \scriptfont\sffam=\elevensf
      \scriptscriptfont\sffam=\elevensf
  \fi
  \def\oldstyle{\fam\@ne\eleveni}%
  \b@ls{13pt}\rm\@xipt%
}
\def\@xipt{}

\def\fourteenpoint{
  \def\rm{\fam0\fourteenrm}%
  \textfont0\fourteenrm  \scriptfont0\tenrm  \scriptscriptfont0\sevenrm%
  \textfont1\fourteeni   \scriptfont1\teni   \scriptscriptfont1\seveni%
  \textfont2\fourteensy  \scriptfont2\tensy  \scriptscriptfont2\sevensy%
  \textfont\itfam=\fourteenit\def\it{\fam\itfam\fourteenit}%
  \ifprod@font
    \scriptfont\itfam=\tenit
      \scriptscriptfont\itfam=\sevenit
  \else
    \scriptfont\itfam=\fourteenit
      \scriptscriptfont\itfam=\fourteenit
  \fi
  \textfont\bffam=\fourteenbf%
    \scriptfont\bffam=\tenbf%
      \scriptscriptfont\bffam=\sevenbf%
  \def\bf{\fam\bffam\fourteenbf}%
  \textfont\slfam=\fourteensl\def\sl{\fam\slfam\fourteensl}%
  \ifprod@font
    \scriptfont\slfam=\tensl
      \scriptscriptfont\slfam=\sevensl
  \else
    \scriptfont\slfam=\fourteensl
      \scriptscriptfont\slfam=\fourteensl
  \fi
  \textfont\ttfam=\fourteentt\def\tt{\fam\ttfam\fourteentt}%
  \ifprod@font
    \scriptfont\ttfam=\tentt
      \scriptscriptfont\ttfam=\seventt
  \else
    \scriptfont\ttfam=\fourteentt
      \scriptscriptfont\ttfam=\fourteentt
  \fi
  \textfont\scfam=\fourteencsc\def\sc{\fam\scfam\fourteencsc}%
  \ifprod@font
    \scriptfont\scfam=\tencsc
      \scriptscriptfont\scfam=\sevencsc
  \else
    \scriptfont\scfam=\fourteencsc
      \scriptscriptfont\scfam=\fourteencsc
  \fi
  \textfont\sffam=\fourteensf\def\sf{\fam\sffam\fourteensf}%
  \ifprod@font
    \scriptfont\sffam=\tensf
      \scriptscriptfont\sffam=\sevensf
  \else
    \scriptfont\sffam=\fourteensf
      \scriptscriptfont\sffam=\fourteensf
  \fi
  \def\oldstyle{\fam\@ne\fourteeni}%
  \b@ls{17pt}\rm\@xivpt%
}
\def\@xivpt{}

\def\seventeenpoint{
  \def\rm{\fam0\seventeenrm}%
  \textfont0\seventeenrm  \scriptfont0\twelverm  \scriptscriptfont0\tenrm%
  \textfont1\seventeeni   \scriptfont1\twelvei   \scriptscriptfont1\teni%
  \textfont2\seventeensy  \scriptfont2\twelvesy  \scriptscriptfont2\tensy%
  \textfont\itfam=\seventeenit\def\it{\fam\itfam\seventeenit}%
  \ifprod@font
    \scriptfont\itfam=\twelveit
      \scriptscriptfont\itfam=\tenit
  \else
    \scriptfont\itfam=\seventeenit
      \scriptscriptfont\itfam=\seventeenit
  \fi
  \textfont\bffam=\seventeenbf%
    \scriptfont\bffam=\twelvebf%
      \scriptscriptfont\bffam=\tenbf%
  \def\bf{\fam\bffam\seventeenbf}%
  \textfont\slfam=\seventeensl\def\sl{\fam\slfam\seventeensl}%
  \ifprod@font
    \scriptfont\slfam=\twelvesl
      \scriptscriptfont\slfam=\tensl
  \else
    \scriptfont\slfam=\seventeensl
      \scriptscriptfont\slfam=\seventeensl
  \fi
  \textfont\ttfam=\seventeentt\def\tt{\fam\ttfam\seventeentt}%
  \ifprod@font
    \scriptfont\ttfam=\twelvett
      \scriptscriptfont\ttfam=\tentt
  \else
    \scriptfont\ttfam=\seventeentt
      \scriptscriptfont\ttfam=\seventeentt
  \fi
  \textfont\scfam=\seventeencsc\def\sc{\fam\scfam\seventeencsc}%
  \ifprod@font
    \scriptfont\scfam=\twelvecsc
      \scriptscriptfont\scfam=\tencsc
  \else
    \scriptfont\scfam=\seventeencsc
      \scriptscriptfont\scfam=\seventeencsc
  \fi
  \textfont\sffam=\seventeensf\def\sf{\fam\sffam\seventeensf}%
  \ifprod@font
    \scriptfont\sffam=\twelvesf
      \scriptscriptfont\sffam=\tensf
  \else
    \scriptfont\sffam=\seventeensf
      \scriptscriptfont\sffam=\seventeensf
  \fi
  \def\oldstyle{\fam\@ne\seventeeni}%
  \b@ls{20pt}\rm\@xviipt%
}
\def\@xviipt{}

\lineskip=1pt      \normallineskip=\lineskip
\lineskiplimit=\z@ \normallineskiplimit=\lineskiplimit


\def\loadboldmathnames{%
  \def\balpha{{\bmath{\alpha}}}%
  \def\bbeta{{\bmath{\beta}}}%
  \def\bgamma{{\bmath{\gamma}}}%
  \def\bdelta{{\bmath{\delta}}}%
  \def\bepsilon{{\bmath{\epsilon}}}%
  \def\bzeta{{\bmath{\zeta}}}%
  \def\boldeta{{\bmath{\eta}}}%
  \def\btheta{{\bmath{\theta}}}%
  \def\biota{{\bmath{\iota}}}%
  \def\bkappa{{\bmath{\kappa}}}%
  \def\blambda{{\bmath{\lambda}}}%
  \def\bmu{{\bmath{\mu}}}%
  \def\bnu{{\bmath{\nu}}}%
  \def\bxi{{\bmath{\xi}}}%
  \def\bpi{{\bmath{\pi}}}%
  \def\brho{{\bmath{\rho}}}%
  \def\bsigma{{\bmath{\sigma}}}%
  \def\btau{{\bmath{\tau}}}%
  \def\bupsilon{{\bmath{\upsilon}}}%
  \def\bphi{{\bmath{\phi}}}%
  \def\bchi{{\bmath{\chi}}}%
  \def\bpsi{{\bmath{\psi}}}%
  \def\bomega{{\bmath{\omega}}}%
  \def\bvarepsilon{{\bmath{\varepsilon}}}%
  \def\bvartheta{{\bmath{\vartheta}}}%
  \def\bvarpi{{\bmath{\varpi}}}%
  \def\bvarrho{{\bmath{\varrho}}}%
  \def\bvarsigma{{\bmath{\varsigma}}}%
  \def\bvarphi{{\bmath{\varphi}}}%
  \def\baleph{{\bmath{\aleph}}}%
  \def\bimath{{\bmath{\imath}}}%
  \def\bjmath{{\bmath{\jmath}}}%
  \def\bell{{\bmath{\ell}}}%
  \def\bwp{{\bmath{\wp}}}%
  \def\bRe{{\bmath{\Re}}}%
  \def\bIm{{\bmath{\Im}}}%
  \def\bpartial{{\bmath{\partial}}}%
  \def\binfty{{\bmath{\infty}}}%
  \def\bprime{{\bmath{\prime}}}%
  \def\bemptyset{{\bmath{\emptyset}}}%
  \def\bnabla{{\bmath{\nabla}}}%
  \def\btop{{\bmath{\top}}}%
  \def\bbot{{\bmath{\bot}}}%
  \def\btriangle{{\bmath{\triangle}}}%
  \def\bforall{{\bmath{\forall}}}%
  \def\bexists{{\bmath{\exists}}}%
  \def\bneg{{\bmath{\neg}}}%
  \def\bflat{{\bmath{\flat}}}%
  \def\bnatural{{\bmath{\natural}}}%
  \def\bsharp{{\bmath{\sharp}}}%
  \def\bclubsuit{{\bmath{\clubsuit}}}%
  \def\bdiamondsuit{{\bmath{\diamondsuit}}}%
  \def\bheartsuit{{\bmath{\heartsuit}}}%
  \def\bspadesuit{{\bmath{\spadesuit}}}%
  \def\bsmallint{{\bmath{\smallint}}}%
  \def\btriangleleft{{\bmath{\triangleleft}}}%
  \def\btriangleright{{\bmath{\triangleright}}}%
  \def\bbigtriangleup{{\bmath{\bigtriangleup}}}%
  \def\bbigtriangledown{{\bmath{\bigtriangledown}}}%
  \def\bwedge{{\bmath{\wedge}}}%
  \def\bvee{{\bmath{\vee}}}%
  \def\bcap{{\bmath{\cap}}}%
  \def\bcup{{\bmath{\cup}}}%
  \def\bddagger{{\bmath{\ddagger}}}%
  \def\bdagger{{\bmath{\dagger}}}%
  \def\bsqcap{{\bmath{\sqcap}}}%
  \def\bsqcup{{\bmath{\sqcup}}}%
  \def\buplus{{\bmath{\uplus}}}%
  \def\bamalg{{\bmath{\amalg}}}%
  \def\bdiamond{{\bmath{\diamond}}}%
  \def\bbullet{{\bmath{\bullet}}}%
  \def\bwr{{\bmath{\wr}}}%
  \def\bdiv{{\bmath{\div}}}%
  \def\bodot{{\bmath{\odot}}}%
  \def\boslash{{\bmath{\oslash}}}%
  \def\botimes{{\bmath{\otimes}}}%
  \def\bominus{{\bmath{\ominus}}}%
  \def\boplus{{\bmath{\oplus}}}%
  \def\bmp{{\bmath{\mp}}}%
  \def\bpm{{\bmath{\pm}}}%
  \def\bcirc{{\bmath{\circ}}}%
  \def\bbigcirc{{\bmath{\bigcirc}}}%
  \def\bsetminus{{\bmath{\setminus}}}%
  \def\bcdot{{\bmath{\cdot}}}%
  \def\bast{{\bmath{\ast}}}%
  \def\btimes{{\bmath{\times}}}%
  \def\bstar{{\bmath{\star}}}%
  \def\bpropto{{\bmath{\propto}}}%
  \def\bsqsubseteq{{\bmath{\sqsubseteq}}}%
  \def\bsqsupseteq{{\bmath{\sqsupseteq}}}%
  \def\bparallel{{\bmath{\parallel}}}%
  \def\bmid{{\bmath{\mid}}}%
  \def\bdashv{{\bmath{\dashv}}}%
  \def\bvdash{{\bmath{\vdash}}}%
  \def\bnearrow{{\bmath{\nearrow}}}%
  \def\bsearrow{{\bmath{\searrow}}}%
  \def\bnwarrow{{\bmath{\nwarrow}}}%
  \def\bswarrow{{\bmath{\swarrow}}}%
  \def\bLeftrightarrow{{\bmath{\Leftrightarrow}}}%
  \def\bLeftarrow{{\bmath{\Leftarrow}}}%
  \def\bRightarrow{{\bmath{\Rightarrow}}}%
  \def\bleq{{\bmath{\leq}}}%
  \def\bgeq{{\bmath{\geq}}}%
  \def\bsucc{{\bmath{\succ}}}%
  \def\bprec{{\bmath{\prec}}}%
  \def\bapprox{{\bmath{\approx}}}%
  \def\bsucceq{{\bmath{\succeq}}}%
  \def\bpreceq{{\bmath{\preceq}}}%
  \def\bsupset{{\bmath{\supset}}}%
  \def\bsubset{{\bmath{\subset}}}%
  \def\bsupseteq{{\bmath{\supseteq}}}%
  \def\bsubseteq{{\bmath{\subseteq}}}%
  \def\bin{{\bmath{\in}}}%
  \def\bni{{\bmath{\ni}}}%
  \def\bgg{{\bmath{\gg}}}%
  \def\bll{{\bmath{\ll}}}%
  \def\bnot{{\bmath{\not}}}%
  \def\bleftrightarrow{{\bmath{\leftrightarrow}}}%
  \def\bleftarrow{{\bmath{\leftarrow}}}%
  \def\brightarrow{{\bmath{\rightarrow}}}%
  \def\bmapstochar{{\bmath{\mapstochar}}}%
  \def\bsim{{\bmath{\sim}}}%
  \def\bsimeq{{\bmath{\simeq}}}%
  \def\bperp{{\bmath{\perp}}}%
  \def\bequiv{{\bmath{\equiv}}}%
  \def\basymp{{\bmath{\asymp}}}%
  \def\bsmile{{\bmath{\smile}}}%
  \def\bfrown{{\bmath{\frown}}}%
  \def\bleftharpoonup{{\bmath{\leftharpoonup}}}%
  \def\bleftharpoondown{{\bmath{\leftharpoondown}}}%
  \def\brightharpoonup{{\bmath{\rightharpoonup}}}%
  \def\brightharpoondown{{\bmath{\rightharpoondown}}}%
  \def\blhook{{\bmath{\lhook}}}%
  \def\brhook{{\bmath{\rhook}}}%
  \def\bldotp{{\bmath{\ldotp}}}%
  \def\bcdotp{{\bmath{\cdotp}}}%
}

\def\,{\relax\ifmmode \mskip\thinmuskip\else \thinspace\fi}
\let\protect=\relax

\long\def\@ifundefined#1#2#3{\expandafter\ifx\csname
  #1\endcsname\relax#2\else#3\fi}




\newtoks\math@groups \math@groups={}
\def\addtom@thgroup#1#2{#1\expandafter{\the#1#2}} 



\def\addtosizeh@ok#1#2#3#4{%
  \expandafter\def\csname @#1pt\endcsname{%
    \def\s@ze{#2}\def\ss@ze{#3}\def\sss@ze{#4}\the\math@groups%
  }%
}



\let\resetsizehook=\addtosizeh@ok


\ifprod@font
  \addtosizeh@ok{viii} {8} {6}  {5}
  \addtosizeh@ok{ix}   {9} {6}  {5}
  \addtosizeh@ok{x}    {10}{7}  {5}
  \addtosizeh@ok{xi}   {11}{8}  {6}
  \addtosizeh@ok{xiv}  {14}{10} {7}
  \addtosizeh@ok{xvii} {17}{12}{10}
\else
  \addtosizeh@ok{viii} {8}     {6}     {5}
  \addtosizeh@ok{ix}   {9}     {6}     {5}
  \addtosizeh@ok{x}    {10}    {7}     {5}
  \addtosizeh@ok{xi}   {10.95} {8}     {6}
  \addtosizeh@ok{xiv}  {14.4}  {10}    {7}
  \addtosizeh@ok{xvii} {17.28} {12}    {10}
\fi

\def\get@font#1#2#3{%
  \edef\fonts@ze{\romannumeral#3}
  \edef\fontn@me{\fonts@ze#1}
  \@ifundefined{\fontn@me}%
    {
     \global\expandafter\font\csname \fontn@me\endcsname=#2 at #3pt}%
    {}%
}

\def\ass@tfont#1#2{%
  \xdef\fam@name{\csname #1\endcsname}%
  \xdef\font@name{\csname #2\endcsname}%
  \let\textfont@name\font@name
  \textfont\fam@name\textfont@name
}

\def\ass@sfont#1#2{%
  \xdef\fam@name{\csname #1\endcsname}%
  \xdef\font@name{\csname #2\endcsname}%
  \let\textfont@name\font@name
  \scriptfont\fam@name\textfont@name
}

\def\ass@ssfont#1#2{%
  \xdef\fam@name{\csname #1\endcsname}%
  \xdef\font@name{\csname #2\endcsname}%
  \let\textfont@name\font@name
  \scriptscriptfont\fam@name\textfont@name
}


\def\NewSymbolFont#1#2{%
  \expandafter\ifx\csname sym#1fam\endcsname\relax 
    \expandafter\newfam\csname sym#1fam\endcsname
    \expandafter\edef\csname sym#1fam\endcsname{\the\allocationnumber}%
    \addtom@thgroup\math@groups{%
      \get@font{#1}{#2}{\s@ze}%
      \ass@tfont{sym#1fam}{\fontn@me}%
      \get@font{#1}{#2}{\ss@ze}%
      \ass@sfont{sym#1fam}{\fontn@me}%
      \get@font{#1}{#2}{\sss@ze}%
      \ass@ssfont{sym#1fam}{\fontn@me}%
    }%
  \else
    \errmessage{Family `#1' already defined}%
  \fi
}


\def\NewMathSymbol#1#2#3#4{%
  \edef\f@mly{\expandafter\hexnumber{\csname sym#3fam\endcsname}}%
  \mathchardef#1="#2\f@mly#4\relax
}


\newif\ifd@f

\def\NewMathDelimiter#1#2#3#4#5#6{%
  \d@ftrue
  \expandafter\ifx\csname sym#3fam\endcsname\relax
    \d@ffalse \errmessage{Family `#3' is not defined}%
  \fi
  \expandafter\ifx\csname sym#5fam\endcsname\relax
    \d@ffalse \errmessage{Family `#5' is not defined}%
  \fi
  \ifd@f
    \edef\f@mly{\expandafter\hexnumber{\csname sym#3fam\endcsname}}%
    \edef\f@mlytw@{\expandafter\hexnumber{\csname sym#5fam\endcsname}}%
    \xdef#1{\delimiter"#2\f@mly #4\f@mlytw@ #6\relax}%
  \fi
}


\def\setboxz@h{\setbox\z@\hbox}
\def\wdz@{\wd\z@}
\def\boxz@{\box\z@}
\def\setbox@ne{\setbox\@ne}
\def\wd@ne{\wd\@ne}

\def\math@atom#1#2{%
   \binrel@{#1}\binrel@@{#2}}
\def\binrel@#1{\setboxz@h{\thinmuskip0mu
  \medmuskip\m@ne mu\thickmuskip\@ne mu$#1\m@th$}%
 \setbox@ne\hbox{\thinmuskip0mu\medmuskip\m@ne mu\thickmuskip
  \@ne mu${}#1{}\m@th$}%
 \setbox\tw@\hbox{\hskip\wd@ne\hskip-\wdz@}}
\def\binrel@@#1{\ifdim\wd2<\z@\mathbin{#1}\else\ifdim\wd\tw@>\z@
 \mathrel{#1}\else{#1}\fi\fi}

\def\m@thit{1}

\def\set@skchar#1{\global\expandafter\skewchar
  \csname\fontn@me\endcsname=#1\relax}

\def\NewMathAlphabet#1#2#3{%
  \def\tst{#3}%
  \ifx\tst\empty\else 
    \expandafter\gdef\csname #1@sc\endcsname{}
  \fi
  \expandafter\def\csname #1\endcsname{
    \protect\csname @#1\endcsname}%
  \expandafter\def\csname @#1\endcsname##1{
    {%
    \begingroup
      \get@font{#1}{#2}{\s@ze}%
      \@ifundefined{#1@sc}{}{\set@skchar{#3}}%
      \ass@tfont{m@thit}{\fontn@me}%
      \get@font{#1}{#2}{\ss@ze}%
      \@ifundefined{#1@sc}{}{\set@skchar{#3}}%
      \ass@sfont{m@thit}{\fontn@me}%
      \get@font{#1}{#2}{\sss@ze}%
      \@ifundefined{#1@sc}{}{\set@skchar{#3}}%
      \ass@ssfont{m@thit}{\fontn@me}%
      \math@atom{##1}{%
      \mathchoice%
        {\hbox{$\m@th\displaystyle##1$}}%
        {\hbox{$\m@th\textstyle##1$}}%
        {\hbox{$\m@th\scriptstyle##1$}}%
        {\hbox{$\m@th\scriptscriptstyle##1$}}}%
    \endgroup
    }%
  }%
}


\newif\iffirstta  \firsttatrue

\def\set@hchar#1{\global\expandafter\hyphenchar
  \csname\fontn@me\endcsname=#1\relax}

\def\NewTextAlphabet#1#2#3{%
  \iffirstta
    \global\firsttafalse
    \newfam\scratchfam
    \edef\scrt@fam{\the\allocationnumber}%
  \fi
  \def\tst{#3}%
  \ifx\tst\empty\else 
    \expandafter\gdef\csname #1@hc\endcsname{}
  \fi
  \expandafter\def\csname #1\endcsname{
    \protect\csname t@#1\endcsname}%
  \long\expandafter\def\csname t@#1\endcsname##1{
    \ifmmode
      \typeout{Warning: do not use \expandafter\string\csname #1\endcsname
        \space in math mode}\fi%
    {%
      \get@font{#1}{#2}{\s@ze}\let\t@xtfnt=\fontn@me\relax
      \@ifundefined{#1@hc}{}{\set@hchar{#3}}%
      \ass@tfont{scrt@fam}{\fontn@me}%
      \get@font{#1}{#2}{\ss@ze}%
      \@ifundefined{#1@hc}{}{\set@hchar{#3}}%
      \ass@sfont{scrt@fam}{\fontn@me}%
      \get@font{#1}{#2}{\sss@ze}%
      \@ifundefined{#1@hc}{}{\set@hchar{#3}}%
      \ass@ssfont{scrt@fam}{\fontn@me}%
      \fam\scratchfam\csname\t@xtfnt\endcsname
    ##1%
    }%
  }%
  \expandafter\def\csname #1shape
    \endcsname{\protect\csname @#1shape\endcsname}%
  \expandafter\def\csname @#1shape\endcsname{
    \ifmmode
      \typeout{Warning: do not use \expandafter\string\csname
        #1shape\endcsname \space in math mode}\fi
      \get@font{#1}{#2}{\s@ze}\let\t@xtfnt=\fontn@me\relax
      \@ifundefined{#1@hc}{}{\set@hchar{#3}}%
      \ass@tfont{scrt@fam}{\fontn@me}%
      \get@font{#1}{#2}{\ss@ze}%
      \@ifundefined{#1@hc}{}{\set@hchar{#3}}%
      \ass@sfont{scrt@fam}{\fontn@me}%
      \get@font{#1}{#2}{\sss@ze}%
      \@ifundefined{#1@hc}{}{\set@hchar{#3}}%
      \ass@ssfont{scrt@fam}{\fontn@me}%
      \fam\scratchfam\csname\t@xtfnt\endcsname
  }%
}


\ifprod@font
  \def\math@itfnt{mtmib10}
  \def\math@syfnt{mtbsy10}
\else
  \def\math@itfnt{cmmib10}
  \def\math@syfnt{cmbsy10}
\fi

\def\m@thsy{2}

\def\bmath{\protect\@bmath}
\def\@bmath#1{%
  {%
  \begingroup
    \get@font{mthit}{\math@itfnt}{\s@ze}\set@skchar{'177}%
    \ass@tfont{m@thit}{\fontn@me}%
    \get@font{mthit}{\math@itfnt}{\ss@ze}\set@skchar{'177}%
    \ass@sfont{m@thit}{\fontn@me}%
    \get@font{mthit}{\math@itfnt}{\sss@ze}\set@skchar{'177}%
    \ass@ssfont{m@thit}{\fontn@me}%
    \get@font{mthsy}{\math@syfnt}{\s@ze}\set@skchar{'60}%
    \ass@tfont{m@thsy}{\fontn@me}%
    \get@font{mthsy}{\math@syfnt}{\ss@ze}\set@skchar{'60}%
    \ass@sfont{m@thsy}{\fontn@me}%
    \get@font{mthsy}{\math@syfnt}{\sss@ze}\set@skchar{'60}%
    \ass@ssfont{m@thsy}{\fontn@me}%
    \math@atom{#1}{%
    \mathchoice%
      {\hbox{$\m@th\displaystyle#1$}}%
      {\hbox{$\m@th\textstyle#1$}}%
      {\hbox{$\m@th\scriptstyle#1$}}%
      {\hbox{$\m@th\scriptscriptstyle#1$}}}%
  \endgroup
  }%
}



\def\sun{\hbox{$\odot$}}

\def\diameter{{\ifmmode\mathchoice
{\ooalign{\hfil\hbox{$\displaystyle/$}\hfil\crcr
{\hbox{$\displaystyle\mathchar"20D$}}}}
{\ooalign{\hfil\hbox{$\textstyle/$}\hfil\crcr
{\hbox{$\textstyle\mathchar"20D$}}}}
{\ooalign{\hfil\hbox{$\scriptstyle/$}\hfil\crcr
{\hbox{$\scriptstyle\mathchar"20D$}}}}
{\ooalign{\hfil\hbox{$\scriptscriptstyle/$}\hfil\crcr
{\hbox{$\scriptscriptstyle\mathchar"20D$}}}}
\else{\ooalign{\hfil/\hfil\crcr\mathhexbox20D}}%
\fi}}

\def\sq{\ifmmode\squareforqed\else{\unskip\nobreak\hfil
\penalty50\hskip1em\null\nobreak\hfil\squareforqed
\parfillskip=0pt\finalhyphendemerits=0\endgraf}\fi}
\def\squareforqed{\hbox{\rlap{$\sqcap$}$\sqcup$}}


\def\bbbc{{\mathchoice {\setbox0=\hbox{$\displaystyle\rm C$}\hbox{\hbox
to0pt{\kern0.4\wd0\vrule height0.9\ht0\hss}\box0}}
{\setbox0=\hbox{$\textstyle\rm C$}\hbox{\hbox
to0pt{\kern0.4\wd0\vrule height0.9\ht0\hss}\box0}}
{\setbox0=\hbox{$\scriptstyle\rm C$}\hbox{\hbox
to0pt{\kern0.4\wd0\vrule height0.9\ht0\hss}\box0}}
{\setbox0=\hbox{$\scriptscriptstyle\rm C$}\hbox{\hbox
to0pt{\kern0.4\wd0\vrule height0.9\ht0\hss}\box0}}}}
\def\bbbq{{\mathchoice {\setbox0=\hbox{$\displaystyle\rm
Q$}\hbox{\raise
0.15\ht0\hbox to0pt{\kern0.4\wd0\vrule height0.8\ht0\hss}\box0}}
{\setbox0=\hbox{$\textstyle\rm Q$}\hbox{\raise
0.15\ht0\hbox to0pt{\kern0.4\wd0\vrule height0.8\ht0\hss}\box0}}
{\setbox0=\hbox{$\scriptstyle\rm Q$}\hbox{\raise
0.15\ht0\hbox to0pt{\kern0.4\wd0\vrule height0.7\ht0\hss}\box0}}
{\setbox0=\hbox{$\scriptscriptstyle\rm Q$}\hbox{\raise
0.15\ht0\hbox to0pt{\kern0.4\wd0\vrule height0.7\ht0\hss}\box0}}}}
\def\bbbt{{\mathchoice {\setbox0=\hbox{$\displaystyle\rm
T$}\hbox{\hbox to0pt{\kern0.3\wd0\vrule height0.9\ht0\hss}\box0}}
{\setbox0=\hbox{$\textstyle\rm T$}\hbox{\hbox
to0pt{\kern0.3\wd0\vrule height0.9\ht0\hss}\box0}}
{\setbox0=\hbox{$\scriptstyle\rm T$}\hbox{\hbox
to0pt{\kern0.3\wd0\vrule height0.9\ht0\hss}\box0}}
{\setbox0=\hbox{$\scriptscriptstyle\rm T$}\hbox{\hbox
to0pt{\kern0.3\wd0\vrule height0.9\ht0\hss}\box0}}}}
\def\bbbs{{\mathchoice
{\setbox0=\hbox{$\displaystyle     \rm S$}\hbox{\raise0.5\ht0\hbox
to0pt{\kern0.35\wd0\vrule height0.45\ht0\hss}\hbox
to0pt{\kern0.55\wd0\vrule height0.5\ht0\hss}\box0}}
{\setbox0=\hbox{$\textstyle        \rm S$}\hbox{\raise0.5\ht0\hbox
to0pt{\kern0.35\wd0\vrule height0.45\ht0\hss}\hbox
to0pt{\kern0.55\wd0\vrule height0.5\ht0\hss}\box0}}
{\setbox0=\hbox{$\scriptstyle      \rm S$}\hbox{\raise0.5\ht0\hbox
to0pt{\kern0.35\wd0\vrule height0.45\ht0\hss}\raise0.05\ht0\hbox
to0pt{\kern0.5\wd0\vrule height0.45\ht0\hss}\box0}}
{\setbox0=\hbox{$\scriptscriptstyle\rm S$}\hbox{\raise0.5\ht0\hbox
to0pt{\kern0.4\wd0\vrule height0.45\ht0\hss}\raise0.05\ht0\hbox
to0pt{\kern0.55\wd0\vrule height0.45\ht0\hss}\box0}}}}
\def\bbbz{{\mathchoice {\hbox{$\sf\textstyle Z\kern-0.4em Z$}}
{\hbox{$\sf\textstyle Z\kern-0.4em Z$}}
{\hbox{$\sf\scriptstyle Z\kern-0.3em Z$}}
{\hbox{$\sf\scriptscriptstyle Z\kern-0.2em Z$}}}}


\def\Nulle{0} 
\def\Afe{1}   
\def\Hae{2}   
\def\Hbe{3}   
\def\Hce{4}   
\def\Hde{5}   


\newcount\LastMac       \LastMac=\Nulle

\newskip\half      \half=5.5pt plus 1.5pt minus 2.25pt
\newskip\one       \one=11pt plus 3pt minus 5.5pt
\newskip\onehalf   \onehalf=16.5pt plus 5.5pt minus 8.25pt
\newskip\two       \two=22pt plus 5.5pt minus 11pt

\def\Half{\addvspace{\half}}
\def\One{\addvspace{\one}}
\def\OneHalf{\addvspace{\onehalf}}
\def\Two{\addvspace{\two}}

\def\Raggedright{
  \rightskip=\z@ plus \hsize\relax
}

\def\Fullout{
  \rightskip=\z@\relax
}

\def\Hang#1#2{
  \hangindent=#1%
  \hangafter=#2\relax
}


\newif\ifsp@page
\def\pagestyle#1{\csname ps@#1\endcsname}
\def\thispagestyle#1{\global\sp@pagetrue\gdef\sp@type{#1}}

\def\ps@titlepage{%
  \def\@oddhead{\eightpoint\noindent \the\CatchLine
    \ifprod@font\else\qquad Printed\ \today\qquad
      (MN plain \TeX\ macros\ v\@version)\fi \hfil}%
  \let\@evenhead=\@oddhead
  \def\@oddfoot{\eightpoint\copyright\ \@pubyear\ RAS\hfil}%
  \def\@evenfoot{\hfil\eightpoint\noindent\copyright\ \@pubyear\ RAS}%
}

\def\ps@headings{%
  \def\@oddhead{\elevenpoint\it\noindent
    \hfill\the\RightHeader\hskip1.5em\rm\folio}%
  \def\@evenhead{\elevenpoint\noindent
    \folio\hskip1.5em\it\the\LeftHeader\hfill}%
  \def\@oddfoot{\eightpoint\noindent\copyright\ \@pubyear\ RAS,
    MNRAS {\bf \@volume}, \@pagerange\hfil}%
  \def\@evenfoot{\hfil\eightpoint\copyright\ \@pubyear\ RAS,
    MNRAS {\bf \@volume}, \@pagerange}%
}

\def\ps@plate{%
  \def\@oddhead{\eightpoint\noindent\plt@cap\hfil}%
  \def\@evenhead{\eightpoint\noindent\plt@cap\hfil}%
  \def\@oddfoot{\eightpoint\noindent\copyright\ \@pubyear\ RAS,
    MNRAS {\bf \@volume}, \@pagerange\hfil}%
  \def\@evenfoot{\hfil\eightpoint\copyright\ \@pubyear\ RAS,
    MNRAS {\bf \@volume}, \@pagerange}%
}



\def\title#1{
  \bgroup
    \vbox to 8pt{\vss}%
    \seventeenpoint
    \Raggedright
    \noindent \strut{\bf #1}\par
  \egroup
}

\def\author#1{
  \bgroup
    \ifnum\LastMac=\Afe \OneHalf\else \vskip 21pt\fi
    \fourteenpoint
    \Raggedright
    \noindent \strut #1\par
    \vskip 3pt%
  \egroup
}

\def\affiliation#1{
  \bgroup
    \vskip -4pt%
    \eightpoint
    \Raggedright
    \noindent \strut {\it #1}\par
  \egroup
  \LastMac=\Afe\relax
}

\def\acceptedline#1{
  \bgroup
    \Two
    \eightpoint
    \Raggedright
    \noindent \strut #1\par
  \egroup
}

\long\def\abstract#1{%
  \bgroup
    \vskip 20pt%
    \leftskip 11pc\rightskip\z@
    \noindent{\ninebf ABSTRACT}\par
    \tenpoint
    \Fullout
    \noindent #1\par
  \egroup
}

\long\def\keywords#1{
  \bgroup
    \Half
    \leftskip 11pc\rightskip\z@
    \tenpoint
    \Fullout
    \noindent\hbox{\bf Key words:}\ #1\par
  \egroup
}


\def\maketitle{%
  \EndOpening
  \ifsinglecol \else \MakePage\fi
}


\def\pageoffset#1#2{\hoffset=#1\relax\voffset=#2\relax}


\def\@nameuse#1{\csname #1\endcsname}
\def\arabic#1{\@arabic{\@nameuse{#1}}}
\def\alph#1{\@alph{\@nameuse{#1}}}
\def\Alph#1{\@Alph{\@nameuse{#1}}}
\def\@arabic#1{\number #1}
\def\@Alph#1{\ifcase#1\or A\or B\or C\or D\else\@Ialph{#1}\fi}
\def\@Ialph#1{\ifcase#1\or \or \or \or \or E\or F\or G\or H\or I\or J\or
   K\or L\or M\or N\or O\or P\or Q\or R\or S\or T\or U\or V\or W\or X\or
   Y\or Z\else\errmessage{Counter out of range}\fi}
\def\@alph#1{\ifcase#1\or a\or b\or c\or d\else\@ialph{#1}\fi}
\def\@ialph#1{\ifcase#1\or \or \or \or \or e\or f\or g\or h\or i\or j\or
   k\or l\or m\or n\or o\or p\or q\or r\or s\or t\or u\or v\or w\or x\or y\or
   z\else\errmessage{Counter out of range}\fi}


\newcount\Eqnno
\newcount\SubEqnno

\def\theeq{\arabic{Eqnno}}
\def\thesubeq{\alph{SubEqnno}}

\def\stepeq{\relax
  \global\SubEqnno \z@
  \global\advance\Eqnno \@ne\relax
  {\rm (\theeq)}%
}

\def\startsubeq{\relax
  \global\SubEqnno \z@
  \global\advance\Eqnno \@ne\relax
  \stepsubeq
}

\def\stepsubeq{\relax
  \global\advance\SubEqnno \@ne\relax
  {\rm (\theeq\thesubeq)}%
}


\newcount\Sec        
\newcount\SecSec
\newcount\SecSecSec

\def\thesection{\arabic{Sec}}
\def\thesubsection{\thesection.\arabic{SecSec}}
\def\thesubsubsection{\thesubsection.\arabic{SecSecSec}}

\Sec=\z@

\def\:{\let\@sptoken= } \:  
\def\:{\@xifnch} \expandafter\def\: {\futurelet\@tempc\@ifnch}

\def\@ifnextchar#1#2#3{%
  \let\@tempMACe #1%
  \def\@tempMACa{#2}%
  \def\@tempMACb{#3}%
  \futurelet \@tempMACc\@ifnch%
}

\def\@ifnch{%
\ifx \@tempMACc \@sptoken%
  \let\@tempMACd\@xifnch%
\else%
  \ifx \@tempMACc \@tempMACe%
    \let\@tempMACd\@tempMACa%
  \else%
    \let\@tempMACd\@tempMACb%
  \fi%
\fi%
\@tempMACd%
}

\def\@ifstar#1#2{\@ifnextchar *{\def\@tempMACa*{#1}\@tempMACa}{#2}}

\newskip\@tempskipb

\def\addvspace#1{%
  \ifvmode\else \endgraf\fi%
  \ifdim\lastskip=\z@%
    \vskip #1\relax%
  \else%
    \@tempskipb#1\relax\@xaddvskip%
  \fi%
}

\def\@xaddvskip{%
  \ifdim\lastskip<\@tempskipb%
    \vskip-\lastskip%
    \vskip\@tempskipb\relax%
  \else%
    \ifdim\@tempskipb<\z@%
      \ifdim\lastskip<\z@ \else%
        \advance\@tempskipb\lastskip%
        \vskip-\lastskip\vskip\@tempskipb%
      \fi%
    \fi%
  \fi%
}

\newskip\@tmpSKIP

\def\addpen#1{%
  \ifvmode
    \if@nobreak
    \else
      \ifdim\lastskip=\z@
        \penalty#1\relax
      \else
        \@tmpSKIP=\lastskip
        \vskip -\lastskip
        \penalty#1\vskip\@tmpSKIP
      \fi
    \fi
  \fi
}

\newcount\@clubpen   \@clubpen=\clubpenalty
\newif\if@nobreak    \@nobreakfalse

\def\@noafterindent{%
  \global\@nobreaktrue
  \everypar{\if@nobreak
              \global\@nobreakfalse
              \clubpenalty \@M
              {\setbox\z@\lastbox}%
              \LastMac=\Nulle\relax%
            \else
              \clubpenalty \@clubpen
              \everypar{}%
            \fi}%
}

\newcount\gds@cbrk   \gds@cbrk=-300

\def\@nohdbrk{\interlinepenalty \@M\relax}

\let\@par=\par
\def\@restorepar{\def\par{\@par}}

\newif\if@endpe   \@endpefalse
 
\def\@doendpe{\@endpetrue \@nobreakfalse \LastMac=\Nulle\relax%
     \def\par{\@restorepar\everypar{}\par\@endpefalse}%
              \everypar{\setbox\z@\lastbox\everypar{}\@endpefalse}%
}

\def\section{\@ifstar{\@ssection}{\@section}}

\def\@section#1{
  \if@nobreak
    \everypar{}%
    \ifnum\LastMac=\Hae \addvspace{\half}\fi
  \else
    \addpen{\gds@cbrk}%
    \addvspace{\two}%
  \fi
  \bgroup
    \ninepoint\bf
    \Raggedright
    \global\advance\Sec \@ne
    \ifappendix
      \global\Eqnno=\z@ \global\SubEqnno=\z@\relax
      \def\ch@ck{#1}%
      \ifx\ch@ck\empty \def\c@lon{}\else\def\c@lon{:}\fi
      \noindent\@nohdbrk APPENDIX\ \thesection\c@lon\hskip 0.5em%
        \uppercase{#1}\par
    \else
      \noindent\@nohdbrk\thesection\hskip 1pc \uppercase{#1}\par
    \fi
    \global\SecSec=\z@
  \egroup
  \nobreak
  \vskip\half
  \nobreak
  \@noafterindent
  \LastMac=\Hae\relax
}

\def\@ssection#1{
  \if@nobreak
    \everypar{}%
    \ifnum\LastMac=\Hae \addvspace{\half}\fi
  \else
    \addpen{\gds@cbrk}%
    \addvspace{\two}%
  \fi
  \bgroup
    \ninepoint\bf
    \Raggedright
    \noindent\@nohdbrk\uppercase{#1}\par
  \egroup
  \nobreak
  \vskip\half
  \nobreak
  \@noafterindent
  \LastMac=\Hae\relax
}

\def\subsection{\@ifstar{\@ssubsection}{\@subsection}}

\def\@subsection#1{
  \if@nobreak
    \everypar{}%
    \ifnum\LastMac=\Hae \addvspace{1pt plus 1pt minus .5pt}\fi
  \else
    \addpen{\gds@cbrk}%
    \addvspace{\onehalf}%
  \fi
  \bgroup
    \ninepoint\bf
    \Raggedright
    \global\advance\SecSec \@ne
    \noindent\@nohdbrk\thesubsection \hskip 1pc\relax #1\par
    \global\SecSecSec=\z@
  \egroup
  \nobreak
  \vskip\half
  \nobreak
  \@noafterindent
  \LastMac=\Hbe\relax
}

\def\@ssubsection#1{
  \if@nobreak
    \everypar{}%
    \ifnum\LastMac=\Hae \addvspace{1pt plus 1pt minus .5pt}\fi
  \else
    \addpen{\gds@cbrk}%
    \addvspace{\onehalf}%
  \fi
  \bgroup
    \ninepoint\bf
    \Raggedright
    \noindent\@nohdbrk #1\par
  \egroup
  \nobreak
  \vskip\half
  \nobreak
  \@noafterindent
  \LastMac=\Hbe\relax
}

\def\subsubsection{\@ifstar{\@ssubsubsection}{\@subsubsection}}

\def\@subsubsection#1{
  \if@nobreak
    \everypar{}%
    \ifnum\LastMac=\Hbe \addvspace{1pt plus 1pt minus .5pt}\fi
  \else
    \addpen{\gds@cbrk}%
    \addvspace{\onehalf}%
  \fi
  \bgroup
    \ninepoint\it
    \Raggedright
    \global\advance\SecSecSec \@ne
    \noindent\@nohdbrk\thesubsubsection \hskip 1pc\relax #1\par
  \egroup
  \nobreak
  \vskip\half
  \nobreak
  \@noafterindent
  \LastMac=\Hce\relax
}

\def\@ssubsubsection#1{
  \if@nobreak
    \everypar{}%
    \ifnum\LastMac=\Hbe \addvspace{1pt plus 1pt minus .5pt}\fi
  \else
    \addpen{\gds@cbrk}%
    \addvspace{\onehalf}%
  \fi
  \bgroup
    \ninepoint\it
    \Raggedright
    \noindent\@nohdbrk #1\par
  \egroup
  \nobreak
  \vskip\half
  \nobreak
  \@noafterindent
  \LastMac=\Hce\relax
}

\def\paragraph#1{
  \if@nobreak
    \everypar{}%
  \else
    \addpen{\gds@cbrk}%
    \addvspace{\one}%
  \fi%
  \bgroup%
    \ninepoint\it
    \noindent #1\ \nobreak%
  \egroup
  \LastMac=\Hde\relax
  \ignorespaces
}


\newif\ifappendix

\def\appendix{%
  \global\appendixtrue
  \def\thesection{\Alph{Sec}}%
  \def\thesubsection{\thesection\arabic{SecSec}}%
  \def\theeq{\thesection\arabic{Eqnno}}%
  \Sec=\z@ \SecSec=\z@ \SecSecSec=\z@ \Eqnno=\z@ \SubEqnno=\z@\relax
}




\def\beginlist{%
  \par\if@nobreak \else\addvspace{\half}\fi%
  \bgroup%
    \ninepoint
    \let\item=\list@item%
}

\def\list@item{%
  \par\noindent\hskip 1em\relax%
  \ignorespaces%
}

\def\endlist{\par\egroup\addvspace{\half}\@doendpe}


\def\beginrefs{%
  \par
  \bgroup
    \eightpoint
    \Fullout
    \let\bibitem=\bib@item
}

\def\bib@item{%
  \par\parindent=1.5em\Hang{1.5em}{1}%
  \everypar={\Hang{1.5em}{1}\ignorespaces}%
  \noindent\ignorespaces
}

\def\endrefs{\par\egroup\@doendpe}


\newtoks\CatchLine

\def\@journal{Mon.\ Not.\ R.\ Astron.\ Soc.\ }  
\def\@pubyear{1994}        
\def\@pagerange{000--000}  
\def\@volume{000}          
\def\@microfiche{}         %

\def\pubyear#1{\gdef\@pubyear{#1}\@makecatchline}
\def\pagerange#1{\gdef\@pagerange{#1}\@makecatchline}
\def\volume#1{\gdef\@volume{#1}\@makecatchline}
\def\microfiche#1{\gdef\@microfiche{and Microfiche\ #1}\@makecatchline}

\def\@makecatchline{%
  \global\CatchLine{%
    {\rm \@journal {\bf \@volume},\ \@pagerange\ (\@pubyear)\ \@microfiche}}%
}

\@makecatchline 

\newtoks\LeftHeader
\def\shortauthor#1{
  \global\LeftHeader{#1}%
}

\newtoks\RightHeader
\def\shorttitle#1{
  \global\RightHeader{#1}%
}

\def\PageHead{
  \begingroup
    \ifsp@page
      \csname ps@\sp@type\endcsname
    \fi
    \ifodd\pageno
      \let\the@head=\@oddhead
    \else
      \let\the@head=\@evenhead
    \fi
    \vbox to \z@{\vskip-22.5\p@%
      \hbox to \PageWidth{\vbox to8.5\p@{}%
        \the@head
      }%
    \vss}%
  \endgroup
  \nointerlineskip
}

\gdef\PageFoot{%
  \nointerlineskip%
  \begingroup
  \ifsp@page
    \csname ps@\sp@type\endcsname
    \global\sp@pagefalse
  \fi
  \vbox to 22pt{\vfil%
    \hbox to \PageWidth{%
      \eightpoint\strut\noindent
      \ifodd\pageno
        \@oddfoot
      \else
        \@evenfoot
      \fi
    }%
  }%
  \endgroup
}

\def\today{%
  \number\day\space
  \ifcase\month\or January\or February\or March\or April\or May\or June\or
    July\or August\or September\or October\or November\or December\fi
  \space\number\year%
}

\def\authorcomment#1{%
  \gdef\PageFoot{%
    \nointerlineskip%
    \vbox to 20pt{\vfil%
      \hbox to \PageWidth{\elevenpoint\noindent \hfil #1 \hfil}}%
  }%
}


\newif\ifplate@page
\newbox\plt@box

\def\beginplatepage{%
  \let\plate=\plate@head
  \let\caption=\fig@caption
  \global\setbox\plt@box=\vbox\bgroup
  \TEMPDIMEN=\PageWidth 
  \hsize=\PageWidth\relax
}

\def\endplatepage{\par\egroup\global\plate@pagetrue}
\def\plate@head#1{\gdef\plt@cap{#1}}


\def\letters{%
  \gdef\folio{\ifnum\pageno<\z@ L\romannumeral-\pageno
    \else L\number\pageno \fi}%
}


\newdimen\mathindent

\global\mathindent=\z@
\global\everydisplay{\global\@dspwd=\displaywidth\displaysetup}


\def\@displaylines#1{
  {}$\displ@y\hbox{\vbox{\halign{$\@lign\hfil\displaystyle##\hfil$\crcr
  #1\crcr}}}${}%
}

\def\@eqalign#1{\null\vcenter{\openup\jot\m@th
  \ialign{\strut\hfil$\displaystyle{##}$&$\displaystyle{{}##}$\hfil
      \crcr#1\crcr}}%
}

\def\@eqalignno#1{
  \global\advance\@dspwd by -\mathindent%
  {}$\displ@y\hbox{\vbox{\halign to\@dspwd%
  {\hfil$\@lign\displaystyle{##}$\tabskip\z@skip
  &$\@lign\displaystyle{{}##}$\hfil\tabskip\centering
  &\llap{$\@lign##$}\tabskip\z@skip\crcr
  #1\crcr}}}${}%
}


\global\let\displaylines=\@displaylines
\global\let\eqalign=\@eqalign
\global\let\eqalignno=\@eqalignno
\global\let\leqalignno=\@eqalignno

\newdimen\@dspwd   \@dspwd=\z@
\newif\if@eqno
\newif\if@leqno
\newtoks\@eqn
\newtoks\@eq

\def\displaysetup#1$${\displaytest#1\eqno\eqno\displaytest}

\def\displaytest#1\eqno#2\eqno#3\displaytest{%
 \if!#3!\ldisplaytest#1\leqno\leqno\ldisplaytest
 \else\@eqnotrue\@leqnofalse\@eqn={#2}\@eq={#1}\fi
 \generaldisplay$$}

\def\ldisplaytest#1\leqno#2\leqno#3\ldisplaytest{%
\@eq={#1}%
 \if!#3!\@eqnofalse\else\@eqnotrue\@leqnotrue
  \@eqn={#2}\fi}

\def\generaldisplay{%
  \if@eqno
    \if@leqno
      \hbox to \displaywidth{\noindent
        \rlap{$\displaystyle\the\@eqn$}%
        \hskip\mathindent$\displaystyle\the\@eq$\hfil}%
    \else
      \hbox to \displaywidth{\noindent
        \hskip\mathindent
        $\displaystyle\the\@eq$\hfil$\displaystyle\the\@eqn$}%
    \fi
  \else
    \hbox to \displaywidth{\noindent
      \hskip\mathindent$\displaystyle\the\@eq$\hfil}%
  \fi
}


\def\@notice{%
  \par\Two%
  \noindent{\b@ls{11pt}\ninerm This paper has been produced using the
    Royal Astronomical Society/Blackwell Science \TeX\ macros.\par}%
}

\outer\def\bye{\@notice\par\vfill\supereject\end}


\def\start@mess{%
  Monthly notices of the RAS journal style (\@typeface)\space
    v\@version,\space \@verdate.%
}

\everyjob{\Warn{\start@mess}}



\newif\if@debug \@debugfalse  

\def\Print#1{\if@debug\immediate\write16{#1}\else \fi}
\def\Warn#1{\immediate\write16{#1}}
\def\wlog#1{}

\newcount\Iteration 

\def\Single{0} \def\Double{1}                 
\def\Figure{0} \def\Table{1}                  

\def\InStack{0}  
\def\InZoneA{1}
\def\InZoneB{2}
\def\InZoneC{3}

\newcount\TEMPCOUNT 
\newdimen\TEMPDIMEN 
\newbox\TEMPBOX     
\newbox\VOIDBOX     

\newcount\LengthOfStack 
\newcount\MaxItems      
\newcount\StackPointer
\newcount\Point         
\newcount\NextFigure    
\newcount\NextTable     
\newcount\NextItem      

\newcount\StatusStack   
\newcount\NumStack      
\newcount\TypeStack     
\newcount\SpanStack     
\newcount\BoxStack      

\newcount\ItemSTATUS    
\newcount\ItemNUMBER    
\newcount\ItemTYPE      
\newcount\ItemSPAN      
\newbox\ItemBOX         
\newdimen\ItemSIZE      

\newdimen\PageHeight    
\newdimen\TextLeading   
\newdimen\Feathering    
\newcount\LinesPerPage  
\newdimen\ColumnWidth   
\newdimen\ColumnGap     
\newdimen\PageWidth     
\newdimen\BodgeHeight   
\newcount\Leading       

\newdimen\ZoneBSize  
\newdimen\TextSize   
\newbox\ZoneABOX     
\newbox\ZoneBBOX     
\newbox\ZoneCBOX     

\newif\ifFirstSingleItem
\newif\ifFirstZoneA
\newif\ifMakePageInComplete
\newif\ifMoreFigures \MoreFiguresfalse 
\newif\ifMoreTables  \MoreTablesfalse  

\newif\ifFigInZoneB 
\newif\ifFigInZoneC 
\newif\ifTabInZoneB 
\newif\ifTabInZoneC

\newif\ifZoneAFullPage

\newbox\MidBOX    
\newbox\LeftBOX
\newbox\RightBOX
\newbox\PageBOX   

\newif\ifLeftCOL  
\LeftCOLtrue

\newdimen\ZoneBAdjust

\newcount\ItemFits
\def\Yes{1}
\def\No{2}


\MaxItems=15
\NextFigure=\z@        
\NextTable=\@ne

\BodgeHeight=6pt
\TextLeading=11pt    
\Leading=11
\Feathering=\z@      
\LinesPerPage=61     
\topskip=\TextLeading
\ColumnWidth=20pc    
\ColumnGap=2pc       

\newskip\ItemSepamount  
\ItemSepamount=\TextLeading plus \TextLeading minus 4pt

\parskip=\z@ plus .1pt
\parindent=18pt
\widowpenalty=\z@
\clubpenalty=10000
\tolerance=1500
\hbadness=1500
\abovedisplayskip=6pt plus 2pt minus 1pt
\belowdisplayskip=6pt plus 2pt minus 1pt
\abovedisplayshortskip=6pt plus 2pt minus 1pt
\belowdisplayshortskip=6pt plus 2pt minus 1pt

\frenchspacing

\ninepoint 

\PageHeight=682pt
\PageWidth=2\ColumnWidth
\advance\PageWidth by \ColumnGap

\pagestyle{headings}




\newcount\DUMMY \StatusStack=\allocationnumber
\newcount\DUMMY \newcount\DUMMY \newcount\DUMMY 
\newcount\DUMMY \newcount\DUMMY \newcount\DUMMY 
\newcount\DUMMY \newcount\DUMMY \newcount\DUMMY
\newcount\DUMMY \newcount\DUMMY \newcount\DUMMY 
\newcount\DUMMY \newcount\DUMMY \newcount\DUMMY

\newcount\DUMMY \NumStack=\allocationnumber
\newcount\DUMMY \newcount\DUMMY \newcount\DUMMY 
\newcount\DUMMY \newcount\DUMMY \newcount\DUMMY 
\newcount\DUMMY \newcount\DUMMY \newcount\DUMMY 
\newcount\DUMMY \newcount\DUMMY \newcount\DUMMY 
\newcount\DUMMY \newcount\DUMMY \newcount\DUMMY

\newcount\DUMMY \TypeStack=\allocationnumber
\newcount\DUMMY \newcount\DUMMY \newcount\DUMMY 
\newcount\DUMMY \newcount\DUMMY \newcount\DUMMY 
\newcount\DUMMY \newcount\DUMMY \newcount\DUMMY 
\newcount\DUMMY \newcount\DUMMY \newcount\DUMMY 
\newcount\DUMMY \newcount\DUMMY \newcount\DUMMY

\newcount\DUMMY \SpanStack=\allocationnumber
\newcount\DUMMY \newcount\DUMMY \newcount\DUMMY 
\newcount\DUMMY \newcount\DUMMY \newcount\DUMMY 
\newcount\DUMMY \newcount\DUMMY \newcount\DUMMY 
\newcount\DUMMY \newcount\DUMMY \newcount\DUMMY 
\newcount\DUMMY \newcount\DUMMY \newcount\DUMMY

\newbox\DUMMY   \BoxStack=\allocationnumber
\newbox\DUMMY   \newbox\DUMMY \newbox\DUMMY 
\newbox\DUMMY   \newbox\DUMMY \newbox\DUMMY 
\newbox\DUMMY   \newbox\DUMMY \newbox\DUMMY 
\newbox\DUMMY   \newbox\DUMMY \newbox\DUMMY 
\newbox\DUMMY   \newbox\DUMMY \newbox\DUMMY

\def\wlog{\immediate\write\m@ne}


\def\GetItemAll#1{%
 \GetItemSTATUS{#1}
 \GetItemNUMBER{#1}
 \GetItemTYPE{#1}
 \GetItemSPAN{#1}
 \GetItemBOX{#1}
}

\def\GetItemSTATUS#1{%
 \Point=\StatusStack
 \advance\Point by #1
 \global\ItemSTATUS=\count\Point
}

\def\GetItemNUMBER#1{%
 \Point=\NumStack
 \advance\Point by #1
 \global\ItemNUMBER=\count\Point
}

\def\GetItemTYPE#1{%
 \Point=\TypeStack
 \advance\Point by #1
 \global\ItemTYPE=\count\Point
}

\def\GetItemSPAN#1{%
 \Point\SpanStack
 \advance\Point by #1
 \global\ItemSPAN=\count\Point
}

\def\GetItemBOX#1{%
 \Point=\BoxStack
 \advance\Point by #1
 \global\setbox\ItemBOX=\vbox{\copy\Point}
 \global\ItemSIZE=\ht\ItemBOX
 \global\advance\ItemSIZE by \dp\ItemBOX
 \TEMPCOUNT=\ItemSIZE
 \divide\TEMPCOUNT by \Leading
 \divide\TEMPCOUNT by 65536
 \advance\TEMPCOUNT \@ne
 \ItemSIZE=\TEMPCOUNT pt
 \global\multiply\ItemSIZE by \Leading
}


\def\JoinStack{%
 \ifnum\LengthOfStack=\MaxItems 
  \Warn{WARNING: Stack is full...some items will be lost!}
 \else
  \Point=\StatusStack
  \advance\Point by \LengthOfStack
  \global\count\Point=\ItemSTATUS
  \Point=\NumStack
  \advance\Point by \LengthOfStack
  \global\count\Point=\ItemNUMBER
  \Point=\TypeStack
  \advance\Point by \LengthOfStack
  \global\count\Point=\ItemTYPE
  \Point\SpanStack
  \advance\Point by \LengthOfStack
  \global\count\Point=\ItemSPAN
  \Point=\BoxStack
  \advance\Point by \LengthOfStack
  \global\setbox\Point=\vbox{\copy\ItemBOX}
  \global\advance\LengthOfStack \@ne
  \ifnum\ItemTYPE=\Figure 
   \global\MoreFigurestrue
  \else
   \global\MoreTablestrue
  \fi
 \fi
}


\def\LeaveStack#1{%
 {\Iteration=#1
 \loop
 \ifnum\Iteration<\LengthOfStack
  \advance\Iteration \@ne
  \GetItemSTATUS{\Iteration}
   \advance\Point by \m@ne
   \global\count\Point=\ItemSTATUS
  \GetItemNUMBER{\Iteration}
   \advance\Point by \m@ne
   \global\count\Point=\ItemNUMBER
  \GetItemTYPE{\Iteration}
   \advance\Point by \m@ne
   \global\count\Point=\ItemTYPE
  \GetItemSPAN{\Iteration}
   \advance\Point by \m@ne
   \global\count\Point=\ItemSPAN
  \GetItemBOX{\Iteration}
   \advance\Point by \m@ne
   \global\setbox\Point=\vbox{\copy\ItemBOX}
 \repeat}
 \global\advance\LengthOfStack by \m@ne
}


\newif\ifStackNotClean

\def\CleanStack{%
 \StackNotCleantrue
 {\Iteration=\z@
  \loop
   \ifStackNotClean
    \GetItemSTATUS{\Iteration}
    \ifnum\ItemSTATUS=\InStack
     \advance\Iteration \@ne
     \else
      \LeaveStack{\Iteration}
    \fi
   \ifnum\LengthOfStack<\Iteration
    \StackNotCleanfalse
   \fi
 \repeat}
}


\def\FindItem#1#2{%
 \global\StackPointer=\m@ne 
 {\Iteration=\z@
  \loop
  \ifnum\Iteration<\LengthOfStack
   \GetItemSTATUS{\Iteration}
   \ifnum\ItemSTATUS=\InStack
    \GetItemTYPE{\Iteration}
    \ifnum\ItemTYPE=#1
     \GetItemNUMBER{\Iteration}
     \ifnum\ItemNUMBER=#2
      \global\StackPointer=\Iteration
      \Iteration=\LengthOfStack 
     \fi
    \fi
   \fi
  \advance\Iteration \@ne
 \repeat}
}


\def\FindNext{%
 \global\StackPointer=\m@ne 
 {\Iteration=\z@
  \loop
  \ifnum\Iteration<\LengthOfStack
   \GetItemSTATUS{\Iteration}
   \ifnum\ItemSTATUS=\InStack
    \GetItemTYPE{\Iteration}
   \ifnum\ItemTYPE=\Figure
    \ifMoreFigures
      \global\NextItem=\Figure
      \global\StackPointer=\Iteration
      \Iteration=\LengthOfStack 
    \fi
   \fi
   \ifnum\ItemTYPE=\Table
    \ifMoreTables
      \global\NextItem=\Table
      \global\StackPointer=\Iteration
      \Iteration=\LengthOfStack 
    \fi
   \fi
  \fi
  \advance\Iteration \@ne
 \repeat}
}


\def\ChangeStatus#1#2{%
 \Point=\StatusStack
 \advance\Point by #1
 \global\count\Point=#2
}



\def\Zone{\InZoneA}

\ZoneBAdjust=\z@

\def\MakePage{
 \global\ZoneBSize=\PageHeight
 \global\TextSize=\ZoneBSize
 \global\ZoneAFullPagefalse
 \global\topskip=\TextLeading
 \MakePageInCompletetrue
 \MoreFigurestrue
 \MoreTablestrue
 \FigInZoneBfalse
 \FigInZoneCfalse
 \TabInZoneBfalse
 \TabInZoneCfalse
 \global\FirstSingleItemtrue
 \global\FirstZoneAtrue
 \global\setbox\ZoneABOX=\box\VOIDBOX
 \global\setbox\ZoneBBOX=\box\VOIDBOX
 \global\setbox\ZoneCBOX=\box\VOIDBOX
 \loop
  \ifMakePageInComplete
 \FindNext
 \ifnum\StackPointer=\m@ne
  \NextItem=\m@ne
  \MoreFiguresfalse
  \MoreTablesfalse
 \fi
 \ifnum\NextItem=\Figure
   \FindItem{\Figure}{\NextFigure}
   \ifnum\StackPointer=\m@ne \global\MoreFiguresfalse
   \else
    \GetItemSPAN{\StackPointer}
    \ifnum\ItemSPAN=\Single \def\Zone{\InZoneB}\relax
     \ifFigInZoneC \global\MoreFiguresfalse\fi
    \else
     \def\Zone{\InZoneA}
     \ifFigInZoneB \def\Zone{\InZoneC}\fi
    \fi
   \fi
   \ifMoreFigures\Print{}\FigureItems\fi
 \fi
\ifnum\NextItem=\Table
   \FindItem{\Table}{\NextTable}
   \ifnum\StackPointer=\m@ne \global\MoreTablesfalse
   \else
    \GetItemSPAN{\StackPointer}
    \ifnum\ItemSPAN=\Single\relax
     \ifTabInZoneC \global\MoreTablesfalse\fi
    \else
     \def\Zone{\InZoneA}
     \ifTabInZoneB \def\Zone{\InZoneC}\fi
    \fi
   \fi
   \ifMoreTables\Print{}\TableItems\fi
 \fi
   \MakePageInCompletefalse 
   \ifMoreFigures\MakePageInCompletetrue\fi
   \ifMoreTables\MakePageInCompletetrue\fi
 \repeat
 \ifZoneAFullPage
  \global\TextSize=\z@
  \global\ZoneBSize=\z@
  \global\vsize=\z@\relax
  \global\topskip=\z@\relax
  \vbox to \z@{\vss}
  \eject
 \else
 \global\advance\ZoneBSize by -\ZoneBAdjust
 \global\vsize=\ZoneBSize
 \global\hsize=\ColumnWidth
 \global\ZoneBAdjust=\z@
 \ifdim\TextSize<23pt
 \Warn{}
 \Warn{* Making column fall short: TextSize=\the\TextSize *}
 \vskip-\lastskip\eject\fi
 \fi
}

\def\MakeRightCol{
 \global\TextSize=\ZoneBSize
 \MakePageInCompletetrue
 \MoreFigurestrue
 \MoreTablestrue
 \global\FirstSingleItemtrue
 \global\setbox\ZoneBBOX=\box\VOIDBOX
 \def\Zone{\InZoneB}
 \loop
  \ifMakePageInComplete
 \FindNext
 \ifnum\StackPointer=\m@ne
  \NextItem=\m@ne
  \MoreFiguresfalse
  \MoreTablesfalse
 \fi
 \ifnum\NextItem=\Figure
   \FindItem{\Figure}{\NextFigure}
   \ifnum\StackPointer=\m@ne \MoreFiguresfalse
   \else
    \GetItemSPAN{\StackPointer}
    \ifnum\ItemSPAN=\Double\relax
     \MoreFiguresfalse\fi
   \fi
   \ifMoreFigures\Print{}\FigureItems\fi
 \fi
 \ifnum\NextItem=\Table
   \FindItem{\Table}{\NextTable}
   \ifnum\StackPointer=\m@ne \MoreTablesfalse
   \else
    \GetItemSPAN{\StackPointer}
    \ifnum\ItemSPAN=\Double\relax
     \MoreTablesfalse\fi
   \fi
   \ifMoreTables\Print{}\TableItems\fi
 \fi
   \MakePageInCompletefalse 
   \ifMoreFigures\MakePageInCompletetrue\fi
   \ifMoreTables\MakePageInCompletetrue\fi
 \repeat
 \ifZoneAFullPage
  \global\TextSize=\z@
  \global\ZoneBSize=\z@
  \global\vsize=\z@\relax
  \global\topskip=\z@\relax
  \vbox to \z@{\vss}
  \eject
 \else
 \global\vsize=\ZoneBSize
 \global\hsize=\ColumnWidth
 \ifdim\TextSize<23pt
 \Warn{}
 \Warn{* Making column fall short: TextSize=\the\TextSize *}
 \vskip-\lastskip\eject\fi
\fi
}

\def\FigureItems{
 \Print{Considering...}
 \ShowItem{\StackPointer}
 \GetItemBOX{\StackPointer} 
 \GetItemSPAN{\StackPointer}
  \CheckFitInZone 
  \ifnum\ItemFits=\Yes
   \ifnum\ItemSPAN=\Single
     \ChangeStatus{\StackPointer}{\InZoneB} 
     \global\FigInZoneBtrue
     \ifFirstSingleItem
      \hbox{}\vskip-\BodgeHeight
     \global\advance\ItemSIZE by \TextLeading
     \fi
     \unvbox\ItemBOX\ItemSep
     \global\FirstSingleItemfalse
     \global\advance\TextSize by -\ItemSIZE
     \global\advance\TextSize by -\TextLeading
   \else
    \ifFirstZoneA
     \global\advance\ItemSIZE by \TextLeading
     \global\FirstZoneAfalse\fi
    \global\advance\TextSize by -\ItemSIZE
    \global\advance\TextSize by -\TextLeading
    \global\advance\ZoneBSize by -\ItemSIZE
    \global\advance\ZoneBSize by -\TextLeading
    \ifFigInZoneB\relax
     \else
     \ifdim\TextSize<3\TextLeading
     \global\ZoneAFullPagetrue
     \fi
    \fi
    \ChangeStatus{\StackPointer}{\Zone}
    \ifnum\Zone=\InZoneC \global\FigInZoneCtrue\fi
  \fi
   \Print{TextSize=\the\TextSize}
   \Print{ZoneBSize=\the\ZoneBSize}
  \global\advance\NextFigure \@ne
   \Print{This figure has been placed.}
  \else
   \Print{No space available for this figure...holding over.}
   \Print{}
   \global\MoreFiguresfalse
  \fi
}

\def\TableItems{
 \Print{Considering...}
 \ShowItem{\StackPointer}
 \GetItemBOX{\StackPointer} 
 \GetItemSPAN{\StackPointer}
  \CheckFitInZone 
  \ifnum\ItemFits=\Yes
   \ifnum\ItemSPAN=\Single
    \ChangeStatus{\StackPointer}{\InZoneB}
     \global\TabInZoneBtrue
     \ifFirstSingleItem
      \hbox{}\vskip-\BodgeHeight
     \global\advance\ItemSIZE by \TextLeading
     \fi
     \unvbox\ItemBOX\ItemSep
     \global\FirstSingleItemfalse
     \global\advance\TextSize by -\ItemSIZE
     \global\advance\TextSize by -\TextLeading
   \else
    \ifFirstZoneA
    \global\advance\ItemSIZE by \TextLeading
    \global\FirstZoneAfalse\fi
    \global\advance\TextSize by -\ItemSIZE
    \global\advance\TextSize by -\TextLeading
    \global\advance\ZoneBSize by -\ItemSIZE
    \global\advance\ZoneBSize by -\TextLeading
    \ifFigInZoneB\relax
     \else
     \ifdim\TextSize<3\TextLeading
     \global\ZoneAFullPagetrue
     \fi
    \fi
    \ChangeStatus{\StackPointer}{\Zone}
    \ifnum\Zone=\InZoneC \global\TabInZoneCtrue\fi
   \fi
  \global\advance\NextTable \@ne
   \Print{This table has been placed.}
  \else
  \Print{No space available for this table...holding over.}
   \Print{}
   \global\MoreTablesfalse
  \fi
}


\def\CheckFitInZone{%
{\advance\TextSize by -\ItemSIZE
 \advance\TextSize by -\TextLeading
 \ifFirstSingleItem
  \advance\TextSize by \TextLeading
 \fi
 \ifnum\Zone=\InZoneA\relax
  \else \advance\TextSize by -\ZoneBAdjust
 \fi
 \ifdim\TextSize<3\TextLeading \global\ItemFits=\No
 \else \global\ItemFits=\Yes\fi}
}

\def\BeginOpening{%
  \ninepoint
  \thispagestyle{titlepage}%
  \global\setbox\ItemBOX=\vbox\bgroup%
    \hsize=\PageWidth%
    \hrule height \z@
    \ifsinglecol\vskip 6pt\fi 
}

\let\begintopmatter=\BeginOpening  

\def\EndOpening{%
  \One
  \egroup
  \ifsinglecol
    \box\ItemBOX%
    \vskip\TextLeading plus 2\TextLeading
    \@noafterindent
  \else
    \ItemNUMBER=\z@%
    \ItemTYPE=\Figure
    \ItemSPAN=\Double
    \ItemSTATUS=\InStack
    \JoinStack
  \fi
}


\newif\if@here  \@herefalse

\def\no@float{\global\@heretrue}
\let\nofloat=\relax 

\def\beginfigure{%
  \@ifstar{\global\@dfloattrue \@bfigure}{\global\@dfloatfalse \@bfigure}%
}

\def\@bfigure#1{%
  \par
  \if@dfloat
    \ItemSPAN=\Double
    \TEMPDIMEN=\PageWidth
  \else
    \ItemSPAN=\Single
    \TEMPDIMEN=\ColumnWidth
  \fi
  \ifsinglecol
    \TEMPDIMEN=\PageWidth
  \else
    \ItemSTATUS=\InStack
    \ItemNUMBER=#1%
    \ItemTYPE=\Figure
  \fi
  \bgroup
    \hsize=\TEMPDIMEN
    \global\setbox\ItemBOX=\vbox\bgroup
      \eightpoint\nostb@ls{10pt}%
      \let\caption=\fig@caption
      \ifsinglecol \let\nofloat=\no@float\fi
}

\def\fig@caption#1{%
  \vskip 5.5pt plus 6pt%
  \bgroup 
    \eightpoint\nostb@ls{10pt}%
    \setbox\TEMPBOX=\hbox{#1}%
    \ifdim\wd\TEMPBOX>\TEMPDIMEN
      \noindent \unhbox\TEMPBOX\par
    \else
      \hbox to \hsize{\hfil\unhbox\TEMPBOX\hfil}%
    \fi
  \egroup
}

\def\endfigure{%
  \par\egroup 
  \egroup
  \ifsinglecol
    \if@here \midinsert\global\@herefalse\else \topinsert\fi
      \unvbox\ItemBOX
    \endinsert
  \else
    \JoinStack
    \Print{Processing source for figure \the\ItemNUMBER}%
  \fi
}


\newbox\tab@cap@box
\def\tab@caption#1{\global\setbox\tab@cap@box=\hbox{#1\par}}

\newtoks\tab@txt@toks
\long\def\tab@txt#1{\global\tab@txt@toks={#1}\global\table@txttrue}

\newif\iftable@txt  \table@txtfalse
\newif\if@dfloat    \@dfloatfalse

\def\begintable{%
  \@ifstar{\global\@dfloattrue \@btable}{\global\@dfloatfalse \@btable}%
}

\def\@btable#1{%
  \par
  \if@dfloat
    \ItemSPAN=\Double
    \TEMPDIMEN=\PageWidth
  \else
    \ItemSPAN=\Single
    \TEMPDIMEN=\ColumnWidth
  \fi
  \ifsinglecol
    \TEMPDIMEN=\PageWidth
  \else
    \ItemSTATUS=\InStack
    \ItemNUMBER=#1%
    \ItemTYPE=\Table
  \fi
  \bgroup
    \eightpoint\nostb@ls{10pt}%
    \global\setbox\ItemBOX=\vbox\bgroup
      \let\caption=\tab@caption
      \let\tabletext=\tab@txt
      \ifsinglecol \let\nofloat=\no@float\fi
}

\def\endtable{%
  \par\egroup 
  \egroup
  \setbox\TEMPBOX=\hbox to \TEMPDIMEN{%
    \eightpoint\nostb@ls{10pt}%
    \hss
    \vbox{%
      \hsize=\wd\ItemBOX
      \ifvoid\tab@cap@box
      \else
        \noindent\unhbox\tab@cap@box
        \vskip 5.5pt plus 6pt%
      \fi
      \box\ItemBOX
      \iftable@txt
        \vskip 10pt%
        \noindent\the\tab@txt@toks
        \global\table@txtfalse
      \fi
    }%
    \hss
  }%
  \ifsinglecol
    \if@here \midinsert\global\@herefalse\else \topinsert\fi
      \box\TEMPBOX
    \endinsert
  \else
    \global\setbox\ItemBOX=\box\TEMPBOX
    \JoinStack
    \Print{Processing source for table \the\ItemNUMBER}%
  \fi
}

\def\UnloadZoneA{%
\FirstZoneAtrue
 \Iteration=\z@
  \loop
   \ifnum\Iteration<\LengthOfStack
    \GetItemSTATUS{\Iteration}
    \ifnum\ItemSTATUS=\InZoneA
     \GetItemBOX{\Iteration}
     \ifFirstZoneA \vbox to \BodgeHeight{\vfil}%
     \FirstZoneAfalse\fi
     \unvbox\ItemBOX\ItemSep
     \LeaveStack{\Iteration}
     \else
     \advance\Iteration \@ne
   \fi
 \repeat
}

\def\UnloadZoneC{%
\Iteration=\z@
  \loop
   \ifnum\Iteration<\LengthOfStack
    \GetItemSTATUS{\Iteration}
    \ifnum\ItemSTATUS=\InZoneC
     \GetItemBOX{\Iteration}
     \ItemSep\unvbox\ItemBOX
     \LeaveStack{\Iteration}
     \else
     \advance\Iteration \@ne
   \fi
 \repeat
}


\def\ShowItem#1{
  {\GetItemAll{#1}
  \Print{\the#1:
  {TYPE=\ifnum\ItemTYPE=\Figure Figure\else Table\fi}
  {NUMBER=\the\ItemNUMBER}
  {SPAN=\ifnum\ItemSPAN=\Single Single\else Double\fi}
  {SIZE=\the\ItemSIZE}}}
}

\def\ShowStack{%
 \Print{}
 \Print{LengthOfStack = \the\LengthOfStack}
 \ifnum\LengthOfStack=\z@ \Print{Stack is empty}\fi
 \Iteration=\z@
 \loop
 \ifnum\Iteration<\LengthOfStack
  \ShowItem{\Iteration}
  \advance\Iteration \@ne
 \repeat
}

\def\B#1#2{%
\hbox{\vrule\kern-0.4pt\vbox to #2{%
\hrule width #1\vfill\hrule}\kern-0.4pt\vrule}
}


\newif\ifsinglecol   \singlecolfalse

\def\onecolumn{%
  \global\output={\singlecoloutput}%
  \global\hsize=\PageWidth
  \global\vsize=\PageHeight
  \global\ColumnWidth=\hsize
  \global\TextLeading=12pt
  \global\Leading=12
  \global\singlecoltrue
  \global\let\onecolumn=\relax
  \global\let\footnote=\sing@footnote
  \global\let\vfootnote=\sing@vfootnote
  \ninepoint 
  \message{(Single column)}%
}

\def\singlecoloutput{%
  \shipout\vbox{\PageHead\vbox to \PageHeight{\pagebody\vss}\PageFoot}%
  \advancepageno
  \ifplate@page
    \shipout\vbox{%
      \sp@pagetrue
      \def\sp@type{plate}%
      \global\plate@pagefalse
      \PageHead\vbox to \PageHeight{\unvbox\plt@box\vfil}\PageFoot%
    }%
    \message{[plate]}%
    \advancepageno
  \fi
  \ifnum\outputpenalty>-\@MM \else\dosupereject\fi%
}

\def\ItemSep{\vskip\ItemSepamount\relax}

\def\ItemSepbreak{\par\ifdim\lastskip<\ItemSepamount
  \removelastskip\penalty-200\ItemSep\fi%
}


\let\@@endinsert=\endinsert 

\def\endinsert{\egroup 
  \if@mid \dimen@\ht\z@ \advance\dimen@\dp\z@ \advance\dimen@12\p@
    \advance\dimen@\pagetotal \advance\dimen@-\pageshrink
    \ifdim\dimen@>\pagegoal\@midfalse\p@gefalse\fi\fi
  \if@mid \ItemSep\box\z@\ItemSepbreak
  \else\insert\topins{\penalty100 
    \splittopskip\z@skip
    \splitmaxdepth\maxdimen \floatingpenalty\z@
    \ifp@ge \dimen@\dp\z@
    \vbox to\vsize{\unvbox\z@\kern-\dimen@}
    \else \box\z@\nobreak\ItemSep\fi}\fi\endgroup%
}


\def\gobbleone#1{}
\def\gobbletwo#1#2{}
\let\footnote=\gobbletwo 
\let\vfootnote=\gobbleone

\def\sing@footnote#1{\let\@sf\empty 
  \ifhmode\edef\@sf{\spacefactor\the\spacefactor}\/\fi
  \hbox{$^{\hbox{\eightpoint #1}}$}\@sf\sing@vfootnote{#1}%
}

\def\sing@vfootnote#1{\insert\footins\bgroup\eightpoint\b@ls{9pt}%
  \interlinepenalty\interfootnotelinepenalty
  \splittopskip\ht\strutbox 
  \splitmaxdepth\dp\strutbox \floatingpenalty\@MM
  \leftskip\z@skip \rightskip\z@skip \spaceskip\z@skip \xspaceskip\z@skip
  \noindent $^{\scriptstyle\hbox{#1}}$\hskip 4pt%
    \footstrut\futurelet\next\fo@t%
}

\def\footnoterule{\kern-3\p@ \hrule height \z@ \kern 3\p@}

\skip\footins=19.5pt plus 12pt minus 1pt
\count\footins=1000
\dimen\footins=\maxdimen

\def\note#1#2{%
  \let\@sf=\empty \ifhmode\edef\@sf{\spacefactor\the\spacefactor}\/\fi
  #1\insert\footins\bgroup
    \eightpoint\b@ls{10pt}\rm
    \interlinepenalty\interfootnotelinepenalty
    \splitmaxdepth\dp\strutbox \floatingpenalty\@MM
    \leftskip\z@skip \rightskip\z@skip \spaceskip\z@skip \xspaceskip\z@skip
    \noindent\footstrut #1$\,$#2\strut\par
  \egroup
  \@sf\relax}


\def\landscape{%
  \global\TEMPDIMEN=\PageWidth
  \global\PageWidth=\PageHeight
  \global\PageHeight=\TEMPDIMEN
  \global\let\landscape=\relax
  \onecolumn
  \message{(landscape)}%
  \raggedbottom
}


\output{%
  \ifLeftCOL
    \global\setbox\LeftBOX=\vbox to \ZoneBSize{\box255\unvbox\ZoneBBOX
      \ifvoid\footins\else
        \vskip\skip\footins\unvbox\footins\fi
    }%
    \global\LeftCOLfalse
    \MakeRightCol
  \else
    \setbox\RightBOX=\vbox to \ZoneBSize{\box255\unvbox\ZoneBBOX
      \ifvoid\footins\else
        \vskip\skip\footins\unvbox\footins\fi
    }%
    \setbox\MidBOX=\hbox{\box\LeftBOX\hskip\ColumnGap\box\RightBOX}%
    \setbox\PageBOX=\vbox to \PageHeight{%
      \UnloadZoneA\box\MidBOX\UnloadZoneC}%
    \shipout\vbox{\PageHead\vbox to \PageHeight{\box\PageBOX\vss}\PageFoot}%
    \advancepageno
    \ifplate@page
      \shipout\vbox{%
        \sp@pagetrue
        \def\sp@type{plate}%
        \global\plate@pagefalse
        \PageHead\vbox to \PageHeight{\unvbox\plt@box\vfil}\PageFoot%
      }%
      \message{[plate]}%
      \advancepageno
    \fi
    \global\LeftCOLtrue
    \CleanStack
    \MakePage
  \fi
}


\Warn{\start@mess}

\newif\ifCUPmtplainloaded 
\ifprod@font
  \global\CUPmtplainloadedtrue
\fi

\def\mnmacrosloaded{} 

\catcode `\@=12 



\fi


\input psfig

\newif\ifAMStwofonts

\ifCUPmtplainloaded \else
  \NewTextAlphabet{textbfit} {cmbxti10} {}
  \NewTextAlphabet{textbfss} {cmssbx10} {}
  \NewMathAlphabet{mathbfit} {cmbxti10} {} 
  \NewMathAlphabet{mathbfss} {cmssbx10} {} 
  \ifAMStwofonts
    \NewSymbolFont{upmath} {eurm10}
    \NewSymbolFont{AMSa} {msam10}
    \NewMathSymbol{\upi}     {0}{upmath}{19}
    \NewMathSymbol{\umu}     {0}{upmath}{16}
    \NewMathSymbol{\upartial}{0}{upmath}{40}
    \NewMathSymbol{\leqslant}{3}{AMSa}{36}
    \NewMathSymbol{\geqslant}{3}{AMSa}{3E}

    \let\leq=\leqslant 
    \let\geq=\geqslant 
  \else
    \def\umu{\mu}
    \def\upi{\pi}
    \def\upartial{\partial}
  \fi
\fi

\pageoffset{-2.5pc}{0pc}

\loadboldmathnames


\pagerange{0--0}    
\pubyear{2000}
\volume{000}

\begintopmatter  

\title{The survival of subhaloes in galaxies and galaxy clusters}
\author{Eelco van Kampen}
\affiliation{Institute for Astronomy, University of Edinburgh, Royal
Observatory, Blackford Hill, Edinburgh EH9 3HJ,\ {\tt evk@roe.ac.uk}}
\shortauthor{E.\ van Kampen}
\shorttitle{Surival of subhaloes}
\acceptedline{Accepted ... Received ...; in original form ...}

\abstract {This paper discusses physical and numerical disruption processes
acting on subhaloes in galaxy haloes and in galaxy cluster haloes,
and compare the effects of these processes on the subhalo abundance
within both types of haloes. N-body simulations with a resolution high
enough not to suffer from overmerging (subhalo disruption due to
numerical processes) show a high abundance of subhaloes in both
galaxies and in galaxy clusters. However, observations seem to show a
high subhalo abundance in galaxy clusters only. Thus, it appears that too
many subhaloes survive in simulated galaxy haloes.

There are five main causes for this apparent galaxy subhalo problem.
The most radical one is a change to a cosmology in which structure formation
is not hierarchical below the galaxy halo mass scale. If this is
unacceptable, four causes remain, of which the most important one appears to be
that dynamical friction is not properly simulated yet, not even for the highest
resolution simulations to date, resulting in an 'undermerging' problem. The 
other causes are (numerical) overmerging, differences in the timing of halo
formation and merging in hierarchical structure formation, and significant
differences is mass-to-light ratios.
The net effect of these four causes is that galaxies have a relatively
low abundance of subhaloes, i.e. dwarfs, while at the same time a large
number of field dwarf galaxies can exist which are dark enough to be missed
observationally.}

\keywords {galaxies: evolution - dark matter - large-scale structure of Universe}
\maketitle   

\section{Introduction}

Hierarchical formation of structure in the universe implies the formation
of galaxy and galaxy cluster haloes through accretion and merging of smaller
entities. These entities usually survive as identifiable subhaloes for some
time. The topic of this paper is the difference in the
fate of subhaloes in galaxy clusters and in galaxies, but also the mismatch
between the high abundance of subhaloes in galaxies simulated for a
hierarchical structure formation model dominated by Cold Dark Matter (e.g.
Moore et al. 1999; Klypin et al. 1999b) and the observed low abundance
of dwarf galaxies (e.g. Mateo 1998).

For the modelling of structure formation one usually employs N-body
simulations, in which the matter distribution is sampled by discrete
particles. Haloes are thus modelled as groups of N-body particles,
which can merge, be disrupted, or survive intact, depending on the
physical processes that are in operation, like dynamical friction,
tidal stripping, and others. However, merging and disruption can
also be due to purely numerical processes, which are especially troublesome
for subhaloes consisting of small numbers of particles (van Kampen 2000
and references theirin).

In order to distinguish between numerical and physical causes for
disruption or survival of subhaloes, one approach is to look
at simulations of the same matter distribution at different resolutions.
If a specific effect is seen at two different resolutions, it is likely
to be physical, whereas if it is only seen to operate for the lower
resolution simulation, it must be a numerical effect.
This test was first performed by van Kampen (1995), but
the simulations were not yet of sufficient resolution to clearly
differentiate between physical and numerical effects, even though
the numerical effects appeared to dominate. Another approach is to set up
experiments where one of the effects is artificially eliminated.
Such tests show that numerical effects dominate for small subhaloes
(van Kampen 2000).

Recently several research groups used simulations with a higher resolution
to look at this problem (Klypin et al. 1999a; Ghinga et al. 1998, 1999;
Moore et al. 1999; Okamoto \& Habe 1999).
However, different group finders and different definitions for disruption
times were used, so a direct comparison of the results is not possible.
Still, the consensus is that increasing the number of particles overcomes,
at least partially, the overmerging problem. Unfortunately, for N-body
simulations on a cosmological scale, this requires the use of
very many particles (on the order of $10^9$), which is not practical.
Furthermore, for the smallest subhaloes the overmerging problem simply remains,
as the disruption timescale depends mainly on the number of particles in the
subhalo.

Now that recent simulations manage to partially resolve the overmerging
problem, a new problem, rather ironically, has surfaced: for some
physical systems {\it too many} subhaloes survive as compared to observed
abundances. Moore et al. (1999) and Klypin et al. (1999b) find that
hierarchical formation models predict many more galaxy subhaloes,
i.e. dwarf subhaloes, than observed.
Thus, the question is whether cosmological initial conditions
are such that few dwarf galaxies form in the first place, or that
they are easily destroyed within our Galaxy {\it and} not replaced.
On a larger scale, clusters of galaxies do contain an abundance of
subhaloes, namely its member galaxies. If these are destroyed
as easily as galaxy subhaloes, then they should be replaced at low
redshift by newly accreted galaxies. Otherwise, the only possibility is
that subhaloes need to be disrupted more easily in galaxy haloes than in
galaxy cluster haloes.

The aim of this paper is to establish what causes the difference in subhalo
abundance between galaxies and galaxy clusters, and whether the survival
or disruption of subhaloes in simulated embedding haloes is effected by
numerical or physical processes. In Section 2 the various numerical and
physical processes acting on subhaloes are listed and discussed, and
quantitative estimates for the corresponding disruption timescales are
given for subhaloes in galaxy cluster and galaxy haloes.
Possible reasons for the discrepancy of the subhalo abundance in
simulations and observations are discussed in Section 3.
Specific problems with the modelling of dynamical friction are examined
in Section 4. The consequences of the timing of hierarchical structure
formation are discussed in Section 5.
Finally, a summary with conclusions is provided in Section 6.

\section{Subhalo disruption processes}

\subsection{Definitions}

Before proceeding to discuss disruption process, it is useful to clarify
the associated terminology.
A {\it halo} is defined as a collapsed and virialized density maximum,
whereas a halo that contains {\it subhaloes} is denoted as an
{\it embedding halo}, or as a {\it parent} halo.
The term {\it overmerging} is used for the numerical processes that
artificially merge haloes and subhaloes in an N-body model by
dissolving the subhalo. The term {\it undermerging} is introduced here to
express the inability to model processes that should cause merging or disruption.
Thus, the term {\it merging} only denotes merging due to physical processes
that actually happen in the numerical simulation used.

\subsection{Numerical disruption processes}

Most galaxies and clusters of galaxies are modelled using N-body
simulations, in which discrete particles sample the true density
distribution. The number of N-body particles sets the resolution
of the simulation. If the resolution is too low, i.e.\ too few
particles are used, discreteness effects become important, even
for {\it softened} particles. Two-body interactions can
{\it evaporate} a halo completely. More importantly, subhaloes
can be dissolved by {\it two-body heating} or {\it tidal heating}.
These two processes are both driven by the particles of the embedding
halo, but tidal heating is driven by encounters between embedding halo
particles and collisionless subhaloes, while two-body heating is
driven by two-body encounters between embedding halo particles and
individual subhalo particles.

The timescales for these processes are expressed in terms of the
{\it crossing time} $t_{\rm c}\equiv r/v$, where $r$ is the
{\it half-mass radius}, and $v$ the {\it typical velocity}
(usually taken to be equal to the velocity dispersion).
The {\it two-body evaporation} timescale for either a halo or a subhalo is 
$$t_{\rm evap}\approx 30 {N\over\ln N} {r\over v}  \eqno\stepeq$$
(e.g. Binney \& Tremaine 1987), where $N$ is the number of particles
in the (sub)halo. The {\it particle-subhalo two-body heating} timescale
is given by (van Kampen 1995, 2000)
$$t_{\rm heat} \approx 0.1 {N_{\rm s} \over \ln( r_{\rm h}/2\epsilon)}
  {r_{\rm h} \over r_{\rm s}}
  {r_{\rm h} \over v_{\rm h}}\ ,  \eqno\stepeq$$
\newcount\twobodyheating
\twobodyheating=\Eqnno
where the subscripts $s$ and $h$ indicate subhalo and embedding halo
respectively. Finally, the timescale for {\it particle-subhalo tidal disruption}
(or {\it impulsive disruption}) is given by van Kampen (2000)
$$t_{\rm tidal} \approx 1.3 N_{\rm s}
  \Bigl({r_{\rm p}\over r_{\rm s}}\Bigr)^{1 \over 2}
  {r_{\rm h} \over r_{\rm s}} {r_{\rm h} \over v_{\rm h}} \eqno\stepeq$$
\newcount\tidalheating
\tidalheating=\Eqnno
for isothermal subhaloes, and
$$t_{\rm tidal} \approx 0.5 N_{\rm s}
  {r_{\rm h} \over r_{\rm s}} {r_{\rm h} \over v_{\rm h}} \eqno\stepeq$$
\newcount\tidalheating
\tidalheating=\Eqnno
for Plummer subhaloes. In both cases, $r_{\rm p}$ is taken to be
equal to the softening length $\epsilon$.
Because both $r_{\rm s}/r_{\rm p}$ and $r_{\rm h}/r_{\rm s}$ are
of order ten, and $\ln(r_{\rm h}/2\epsilon)$ is around 5,
the two disruption times are approximately $0.2 N_{\rm s}$ and
$5 N_{\rm s}$ embedding halo crossing times respectively.
Both timescales are smaller than the evaporation timescale, but
particle-subhalo two-body heating is clearly the dominant effect.
Derivations and a more detailed discussion are published elsewhere
(van Kampen 2000).

%

Softening alleviates the problem of two-body effects somewhat, but
softened particle groups are more extended and less strongly bound
(van Kampen 1995). This re-enhances two-body disruption processes,
which are more efficient for more extended subhaloes, as shown below.
More importantly, softening effects the timescales for some of the
{\it physical} disruption processes as well, especially those that
depend on the subhalo size.

\subsection{Physical disruption processes}

\subsubsection{Mean tidal field of embedding halo}

One obvious physical mechanism operating on subhaloes
is heating, stripping, or even disruption by the mean tidal field of the
embedding halo. Various estimates and tests for the timescale and
associated tidal radius exist in the literature
(e.g.\ Allen \& Richstone 1988; Heisler \& White 1990;
van Kampen 1995; Moore et al.\ 1996; Klypin et al.\ 1999).
Unfortunately, the tidal disruption timescale sensitively depends on
the density profile of the embedding halo and the actual orbit
of the subhalo.
Tidal disruption is most efficient on circular orbits near the
core radius of the embedding halo, but less so on radial orbits
where the subhalo spends most of its time outside the `danger zone',
a broad shell around the core where the tidal limit is smallest. Such
subhaloes lose mass due to tidal shocks which are less efficient than a
constant tidal force along a circular orbit (e.g.\ Moore et al.\ 1996).

Thus, it is unlikely that the mean tidal field completely destroys a
subhalo, usually the subhalo is only stripped down to its tidal radius.
If the mean tidal field destroys subhaloes, this should have been
seen in high-resolution simulations.
However, most authors find that a large fraction of
subhaloes survive as identifyable entities, even though they can
suffer significant mass loss and are tidally limited.

As pointed out by van Kampen (1995), groups of softened N-body particles
are more extended and less bound than physical haloes,
and are therefore not only more vulnerable to the mean tidal field,
but also to two-body heating. Furthermore, both processes work in the same
direction and therefore accelerate each other.
This complicates the study of tidal processes and their effects on the
evolution of subhaloes. One solution is to take a fixed potential for the
embedding halo, and only use particles for the subhalo
(e.g.\ Heisler \& White 1990; Moore et al.\ 1996; van Kampen 2000).
The alternative solution is to increase the resolution of the simulation;
if a sufficient number of particles is employed, the subhalo will not be
destroyed by numerical processes, but tidal processes will still be
artificially enhanced.

\subsubsection{Subhalo-subhalo tidal heating}

N-body particles are relatively massive, and therefore
tidally heat subhaloes (Moore et al.\ 1996; van Kampen 2000), a
process which is independent of the number of particles in the subhalo,
as subhaloes are assumed to be collisionless for this process. 
Just like N-body particles can tidally disturb subhaloes,
subhaloes also tidally disturbed each other.
Moore et al. (1996) estimate a timescale for this process by scaling
the timescale for tidal heating of subhaloes by individual N-body
particles. However, there
are two problems with their estimate: their particle-subhalo
timescale estimate is incorrect, and their scaling to subhalo-subhalo
tidal heating misses a term.

The first problem is that the estimate for the particle-subhalo tidal
heating timescale is longer than the estimate Moore et al. give by a
factor of $13 (r_{\rm s}/r_{\rm p})^{3/2}$ for isothermal subhaloes,
or $5 (r_{\rm s}/\epsilon)^2$ for Plummer subhaloes (see van Kampen 2000).
For isothermal subhaloes, the corrected timescale estimate is
$$t_{\rm dis} \approx
  1100 \Bigl({v_{\rm h}\over 1700\ {\rm km\ s}^{-1}}\Bigr)
  \Bigl({\epsilon^{1\over 4}r^{3\over 4}_{\rm s}\over 10\ {\rm kpc}}\Bigr)^2
  \Bigl({10^9 {\rm M}_{\sun} \over m_{\rm p}}\Bigr) {\rm Gyr}\ .
  \eqno\stepeq$$
\newcount\revisedMooreetal
\revisedMooreetal=\Eqnno
Moore et al. also assumed all subhaloes to be tidally truncated, 
which for isothermal haloes means
$r_{\rm s}\approx r_{\rm h}v_{\rm s}/(3v_{\rm h})$.
If we do not assume subhaloes to be tidally truncated,
the corrected disruption timescale is given by eq. (\the\tidalheating).
In both cases $r_{\rm p}$ is set to the softening length $\epsilon$.

The second problem is the relative scaling to subhalo-subhalo heating:
Moore et al. state that the rate of energy input via impulse encounters
scales as $m^2_{\rm p} n_{\rm p}$, and that therefore the relative
importance of subhalo-subhalo encounters versus subhalo-particle
encounters can be written as $f m_{\rm s}/m_{\rm p}$, where $f$ is the
fraction of mass in subhaloes. However, this statement is incorrect,
because for the tidal approximation that Moore et al. (1996) adopt,
the rate of energy input also depends on the size of the
perturber, as eq. (3) of Moore et al. (1996) clearly shows.
So, the rate of energy input actually scales as
$m^2_{\rm p} n_{\rm p} / r^2_{\rm p}$, and the relative importance
therefore as $f (m_{\rm s} / m_{\rm p}) (r_{\rm p}/r_{\rm s})^2$.
This would imply that subhalo-subhalo tidal heating is actually
{\it less} important than particle-subhalo tidal heating,
contrary to the claim of Moore et al. (1996).
However, the tidal approximation cannot be extrapolated down to
$r_{\rm p}=\epsilon$, as Moore et al. (1996) do. Furthermore,
penetrating encounters need to be taken into account as well,
so that the energy change is given by a smooth interpolation between
penetrating and distant encounters (see van Kampen 2000 for details).
For isothermal subhaloes, this means that the energy input
scales as $m^2_{\rm p} n_{\rm p} r^{-1/2}_{\rm p} r^{-3/2}_{\rm s}$
(van Kampen 2000).
Taking this into account, subhalo-subhalo tidal heating is 
$f (m_{\rm s} / m_{\rm p}) (r_{\rm p}/r_{\rm s})^{1\over 2}$ times
faster than particle-subhalo tidal heating. With $f m_{\rm s} / m_{\rm p}$
of order ten, this amounts to a factor of a few.

We thus find that the subhalo-subhalo heating timescale is a factor
$13 (r_{\rm s}/r_{\rm p})^2$ larger than given by Moore et al. (1996),
which is more than an order of magnitude.
With $m_{\rm s}/m_{\rm p}=N_{\rm s}$,
the subhalo-subhalo tidal heating timescale is
$$t_{\rm dis,ss} \approx {2.4\over f} {r_{\rm h} \over r_{\rm s}}
  {r_{\rm h} \over v_{\rm h}}\ . \eqno\stepeq$$
Typically, $r_{\rm h}/r_{\rm s}$ is of order ten, and $f\approx 1/4$,
so the disruption timescale is at least $100$ embedding halo crossing times,
which for cluster galaxy haloes is around 0.3 Gyr, and the disruption
timescale is at least 30 Gyr.

However, if one assumes tidal truncation for the subhaloes,
as Moore et al. (1996) do, one gets
$$t_{\rm dis,ss} \approx
  {7.2 \over f} {v_{\rm h} \over v_{\rm s}} {r_{\rm h} \over v_{\rm h}} =
  {72 \over f} \Bigl({r_{\rm h} \over 1\ {\rm Mpc}}\Bigr)
  \Bigl({100\ {\rm km\ s}^{-1} \over v_{\rm s}}\Bigr) {\rm Gyr}\ .
  \eqno\stepeq$$
This estimate again is more than an order of magnitude larger than
the estimate given by Moore et al. (1996, their eq. 5 at
$R_{\rm c}=r_{\rm h}$).
Tidal truncation of sunhaloes clearly reduces the effectiveness of the
subhalo-subhalo tidal disruption process by a factor of a few.
Still, it remains unclear whether the assumption of tidal truncation is
justified, as it takes time for the tidal truncation to take place, if it
happens at all. When a halo becomes a subhalo, it will certainly not be
truncated straight away.

Note that the timescales as derived above are for unchanging perturbers.
However, as the perturbers get disrupted by each other, or by any
other means, $f$ will decrease, and the disruption timescale will increase,
thus slowing down the subhalo-subhalo tidal heating process.
Indeed, Klypin et al. (1999) make the interesting point that even though
subhalo-subhalo tidal heating seems like an important effect, in practice
many of the subhaloes, especially the smaller ones, get disrupted before
they can participate in this process for a long enough time.
This also means that at any one time there are not many subhaloes left to
tidally heat each other, but there will be an increasing amount of loose
N-body particles in the embedding halo to drive two-body heating.

In concluding, subhalo-subhalo tidal heating is clearly not an important
physical process within galaxies or galaxy clusters, whether subhaloes
are tidally truncated or not.


\subsubsection{Dynamical friction}

The physical mechanism of dynamical friction leads to the complete
destruction of subhaloes by bringing them to the centre of their
embedding halo, where they merge with the core region.
This process is well-known and extensively discussed in the literature
(e.g. Saslaw 1985; Binney \& Tremaine 1987; and references therein).

The actual orbit of the subhalo is important, as for the mean
tidal field. An estimate for the dynamical friction timescale of
a subhalo at $r_{\rm h}$ on a nearly circular orbit within an isothermal
embedding halo is
$$t_{\rm fric}\approx {1.2\over \ln\Lambda}
   {r^2_{\rm h} v_{\rm h}\over G m_{\rm s}}
   \approx {0.5\over \ln\Lambda}
   {m_{\rm h}\over m_{\rm s}} {r_{\rm h}\over v_{\rm h}}\
   \eqno\stepeq$$
\newcount\dynfric
\dynfric=\Eqnno
(Binney \& Tremaine 1987, their eq. 7-26),
where we have used the virial theorem for the embedding halo,
and $\ln\Lambda$ is the Coulomb logarithm, with
$\Lambda\approx m_{\rm h}/m_{\rm s}$ for point masses,
and
$$\ln\Lambda={1\over m^2_{\rm s}} \int_0^{b_{\rm max}} b^3
  \Bigl[\int_b^\infty{m_{\rm s}(r) {\rm d}r\over r^2 (r^2-b^2)^{1/2}}\Bigr]^2
  {\rm d}b \eqno\stepeq$$
for extended subhaloes (White 1976). The deceleration by
dynamical friction is given by
$$\dot{\bf v} = -4\pi\ln\Lambda G^2 m_{\rm s}\rho(<v)
  {{\bf v}\over v^3}\ , \eqno\stepeq$$
\newcount\dynfricterm
\dynfricterm=\Eqnno
where $\rho(<v)$ is the density of background particles moving slower
than the subhalo.
For a Maxwellian velocity distribution with dispersion $\sigma$ this
is given by (Binney \& Tremaine 1987)
$$\rho(<v) = \rho({\bf r})
  \Bigl[{\rm erf}\Bigl({v\over\sqrt{2}\sigma}\Bigr) -
  \sqrt{2\over\pi}{v\over\sigma} e^{-v^2/2\sigma^2}\Bigr]\ .
  \eqno\stepeq $$

The most massive subhaloes are destroyed most rapidly, and a maximum mass
for a subhalo can be set by setting the dynamical friction timescale to
the age of the embedding halo.
The dynamical friction timescale is generally {\it shorter} for
eccentric orbits, by a factor of $\eta^{0.53}$ (van den Bosch et al. 1999),
where $\eta$ is the orbital circularity, defined as the dimensionless
fraction $J/J_{\rm c}$, with $J_{\rm c}$ the angular momentum of a circular
orbit.

For both galaxies and galaxy clusters the mean crossing time is around
0.3 Gyr (with significant scatter), so the dynamical friction
timescale will be around $0.04 m_{\rm h}/ m_{\rm s}$ Gyr (eq. \the\dynfric),
where we have used that $\eta\approx 0.6$ (median value, van den Bosch et al.
1999), and that the Coulomb logarithm $\ln \Lambda$ is of order 3.

If the mass of our Galaxy is $\approx 10^{12}$ M$_{\sun}$, most
subhaloes with $m_{\rm h}>5\times 10^9$ M$_\odot$ should have been
destroyed over the lifetime of our Galaxy. This implies that subhaloes
that are observed today have entered the halo only recently.
A fraction of the subhaloes with masses of around $10^9$ M$_{\sun}$
or less might also have been destroyed, very much depending on their orbits.

For a rich galaxy cluster with mass $10^{15}$ M$_\odot$, galaxy
haloes with masses over $10^{13}$ M$_\odot$ are likely to be destroyed
within the lifetime of a galaxy cluster.
However, as clusters of galaxies are relatively young objects,
secondary infall of new galaxy haloes is ongoing, and destroyed haloes
can be replaced. This makes it hard to
test estimates for the dynamical fraction timescale observationally.

Tormen, Diaferio \& Syer (1998) tested dynamical friction within galaxy
clusters numerically, but unfortunately their simulations contain just
20.000 particles for the embedding halo, which means that many of their
subhaloes disrupt through particle-subhalo two-body heating.
The simulations of Ghigna et al. (1998) employ a hundred
times more particles, and therefore a fair number of subhaloes survive in
their simulations. They do not address dynamical friction, but do state that
the subhaloes are tracers of their embedding halo. In other words, dynamical
friction does not seem to operate in their galaxy cluster simulation.
We discuss this further in Section 4.

\section{The difference between galaxy and cluster subhaloes}

While it is obvious that subhaloes in galaxy cluster haloes exist
(by definition), this is much less obvious for galaxy subhaloes.
Our Galaxy contains just two subhaloes with a significant mass
(the Magellanic Clouds, with a mass of $\approx 10^{10}$ M$_\odot$
for the LMC), while the remaining subhaloes found are much less massive
(eg.\ Mateo 1998).
Clearly, galaxy subhaloes are either destroyed by a physical mechanism,
are much darker (i.e. have a much larger mass-to-light ratio) than cluster
galaxies, or were never formed in large numbers in the first place,
for example if the power spectrum of density fluctuations turns over
below or near the galaxy mass-scale. There are five main explanations of
this sort, some of which have already been proposed in the literature

\paragraph{(1) The initial density fluctuation spectrum lacks power
on small scales}

This is the most radical explanation, and has major consequences
for cosmology. Moore et al.\ (1999) look at standard
CDM only, whereas Klypin et al.\ (1999a) consider both standard CDM and
$\Lambda$CDM, the latter being the favoured model at the moment 
(e.g.\ Efstathiou 2000b). Thus, the problem could only exist for CDM-like
cosmologies. Indeed, a range of models that are
distinctly different from CDM-like models have been proposed:
Warm Dark Matter (e.g. Bardeen et al. 1986) and self-interacting dark matter
(e.g. Spergel \& Steinhardt 1999) are two examples, or even a combination
of the two (Hannestad \& Scherrer 2000). The self-interacting
dark matter cosmology is the most tested recently, but seems a mixed blessing
(e.g. Moore et al. 2000), as it does not actually solve the galaxy subhalo
abundance problem (Yoshida et al. 2000).
The Warm Dark Matter cosmology looks the more promising alternative,
as it provides a solution to a problem with the angular momentum of
disk galaxies as well (Sommer-Larsen \& Dolgov 2000).

\paragraph{(2) Dwarf galaxies are much darker than galaxies}

In hierarchical galaxy formation scenarios there usually is a significant
difference in mass-to-light ratio between dwarf galaxies and `normal' galaxies,
in the sense that the former are much darker than the latter. This is due to
the necessity of strong `feedback', a mechanism in which supernovae reheat
a significant fraction of cold gas in the halo to such a temperature that
star formation stops for that fraction of gas.
In hierarchical galaxy formation some form of feedback is necessary to
prevent turning the majority of baryons into stars at high redshift
(e.g. Efstathiou 2000a and references theirin), but the strength
needed is quite uncertain.

However, even if dwarfs are much darker than galaxies, a large abundance
of dark subhaloes within galaxies might still be problem with respect to the
survival of disks (T\'oth \& Ostriker 1992).
Therefore, the dwarf haloes within galaxies should not survive as long
and in such high abundances as the simulations seem to indicate.

\paragraph{(3) Overmerging produces too many small haloes}

Overmerging always plagues N-body haloes and subhaloes consisting of
less than a hundred particles or so, as discussed in Section 2.2.
Overmerging effects the smallest haloes in the simulations
of Klypin et al. (1999b) and Moore et al. (1999), which have particles
masses of $4\times 10^6 h^{-1}$ M$_{\sun}$ (for the whole simulation volume)
and $10^6 h^{-1}$ M$_{\sun}$ (only within a sphere of twice the virial
radius) respectively. Thus, many subhaloes with masses below $10^8$ M$_{\sun}$
are destroyed through overmerging, which means that too many haloes
with masses over $10^7$ M$_{\sun}$ are identified as single (dwarf) galaxy
haloes, while they should contain several even smaller (dwarf) galaxies and 
a population of globular clusters.

\paragraph{(4) Subhaloes survive in simulations only because of numerical
limitations}

The N-body simulation method is an approximate method, so there could
be a numerical problem which causes the mismatch with observations.
One possible problem is the inability to numerically simulate dynamical
friction, even at the highest resolution achieved to date.
In other words, there might be a numerical `undermerging' problem:
not enough subhaloes are destroyed within galaxy haloes in the simulation.
We explore this possibility in Section 4.

\paragraph{(5) Galaxy clusters contain more subhaloes than galaxies
due to the timing of hierarchical structure formation}

The timing of halo formation and merging within the hierarchical formation
of structure in the Universe can cause significant differences in
the destruction rate of subhaloes in embedding haloes of different mass,
but also differences in the rate at which destroyed subhaloes are
replaced by newly accreted ones.
Whatever process causes subhalo destruction,
any such process is given more time to operate within galaxy haloes,
as these form earlier than galaxy cluster haloes. Furthermore,
galaxy clusters clearly show active
secondary infall, whereas this is not obvious at all for galaxies.
Thus, even if destruction is as effective in clusters as in galaxies,
one still sees more subhaloes in clusters because destroyed subhaloes are
being replaced, whereas galaxy subhaloes are not.

\section{Dynamical friction revisited}

In the derivation of the dynamical friction timescale
(e.g. Binney \& Tremaine 1987) it is assumed that the subhalo moves through
a homogeneous background, or at least through a sea of a large number of
small particles. The highest resolution
simulations (e.g. Moore et al. 1999 and Klypin et al. 1999a) have just
over $10^6$ particles. Therefore, it is likely that the numerical resolution
is not good enough to properly sample the gravitational wake that provides
the effective drag force acting on a subhalo moving through an embedding halo
(e.g. Mulder 1983), which contains an almost infinite abundance of dark
matter particles.
Zaritsky and White (1988) already concluded that simulations of dynamical
friction sensitively depend on subtle details of the simulation technique,
so resolution is likely to be an important factor.

The possibility of `undermerging', due to the inability to properly
simulate dynamical friction, can be tested using simulations of the same
halo-subhalo system at very different resolutions, i.e. particle numbers.
Previously, such test have only been performed for a small range in particle
number (e.g. Cora, Muzzio \& Vergne 1997; van den Bosch et al. 1999).
Here, we discuss series of simulations of equilibrium
Plummer models, where for each series we ran four simulations, with $10^3$,
$10^4$, $10^5$, and $10^6$ particles for the embedding halo. The $10^6$
particle simulations are the largest to date performed for this purpose.

\subsection{Global and local dynamical friction}

In the original derivation of the dynamical friction formula by
Chandrasekhar (1943) the object under study moves through an infinite, uniform
medium, and dynamical friction is solely driven by the density response.
However, a subhalo orbiting within an embedding halo will also tidally
deform that halo, and such global distortions exerts a torque on the
subhalo, thus changing its angular momentum. We call this effect
{\it global} dynamical friction, to contrast it with the {\it local}
dynamical friction driven by the graviational wake (density response)
of the subhalo.
There has been much debate in the literature over whether dynamical friction
is predominantly local or global (e.g. Zaritsky \& White 1988;
Weinberg 1989; Cora et al. 1997; Colpi et al. 1999). Most authors conclude
that it is local, on the basis that timescales measured in simulations
are in fair agreement with Chandrasekhar's formula, even though that was
not derived for a halo-subhalo configuration.
However, these tests are generally for a {\it single} subhalo on a
circular orbit, which is optimal for global dynamical friction. Thus, in
such simulations the wake might not be modelled properly at all, while
the subhalo still decays due to global tidal distortions, which
is not incorporated in the formalism of Chandrasekhar (1943).

In reality, {\it many} subhaloes populate the embedding halo at any one time,
on a variety of orbits. This means that global distortions induced by
each of these subhaloes add up to form a stochastic net distortion pattern.
Thus, the angular momentum changes stochastically as well, and the cumulative
effect on the subhalo is zero, i.e. global dynamical friction becomes
insignificant. Therefore, subhaloes can thus only decay through local dynamical
friction, but only if the wake is properly modelled in the N-body simulation.


\beginfigure{1}
{\vskip 0.2cm \psfig{file=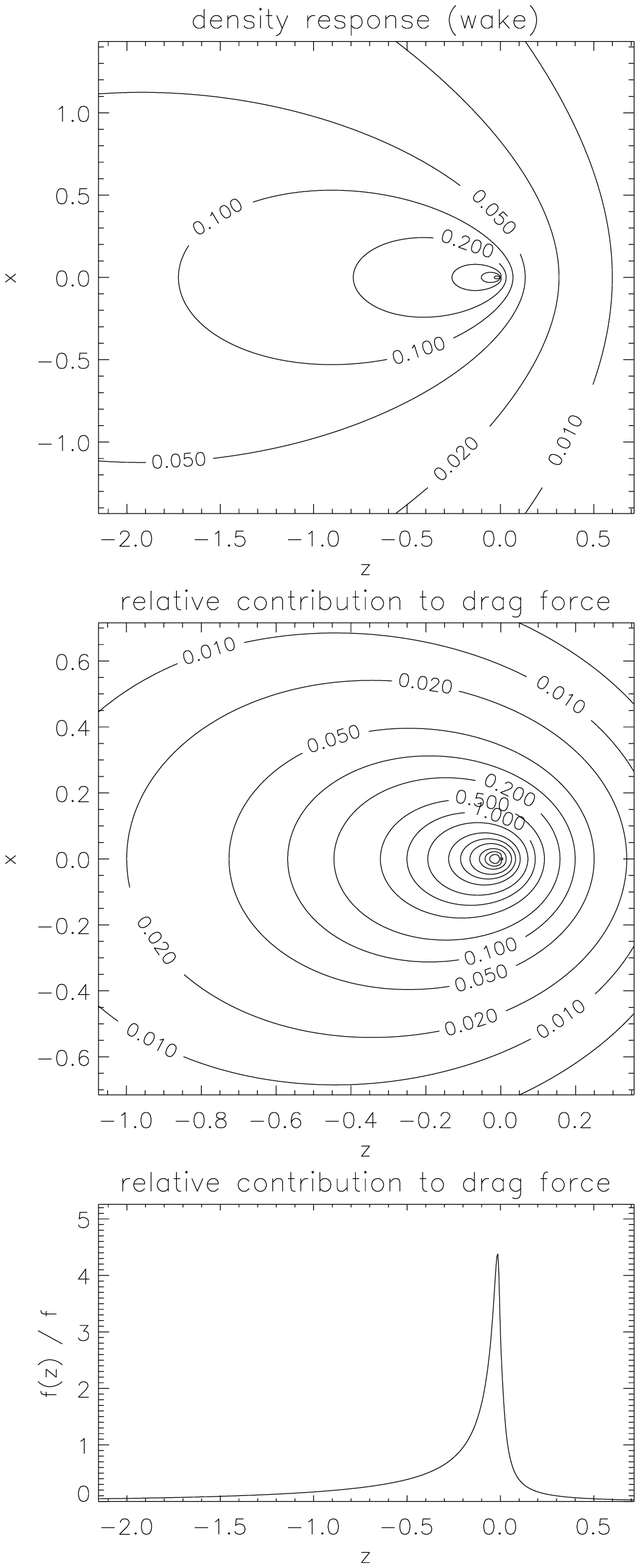,width=9.0cm,silent=1} \vskip -2.0cm}
\caption{{\bf Figure 1a.} The top panel shows the density response, or
gravitational wake), that drives dynamical friction for a compact subhalo
($R_{\rm h}/R_{\rm s}=6$). The middle panel shows the relative contribution
to the total drag force on the subhalo as a function of position. The bottom
panel shows the same, but as a function of $z$ only.}
\endfigure

\beginfigure{2}
{\vskip 0.2cm \psfig{file=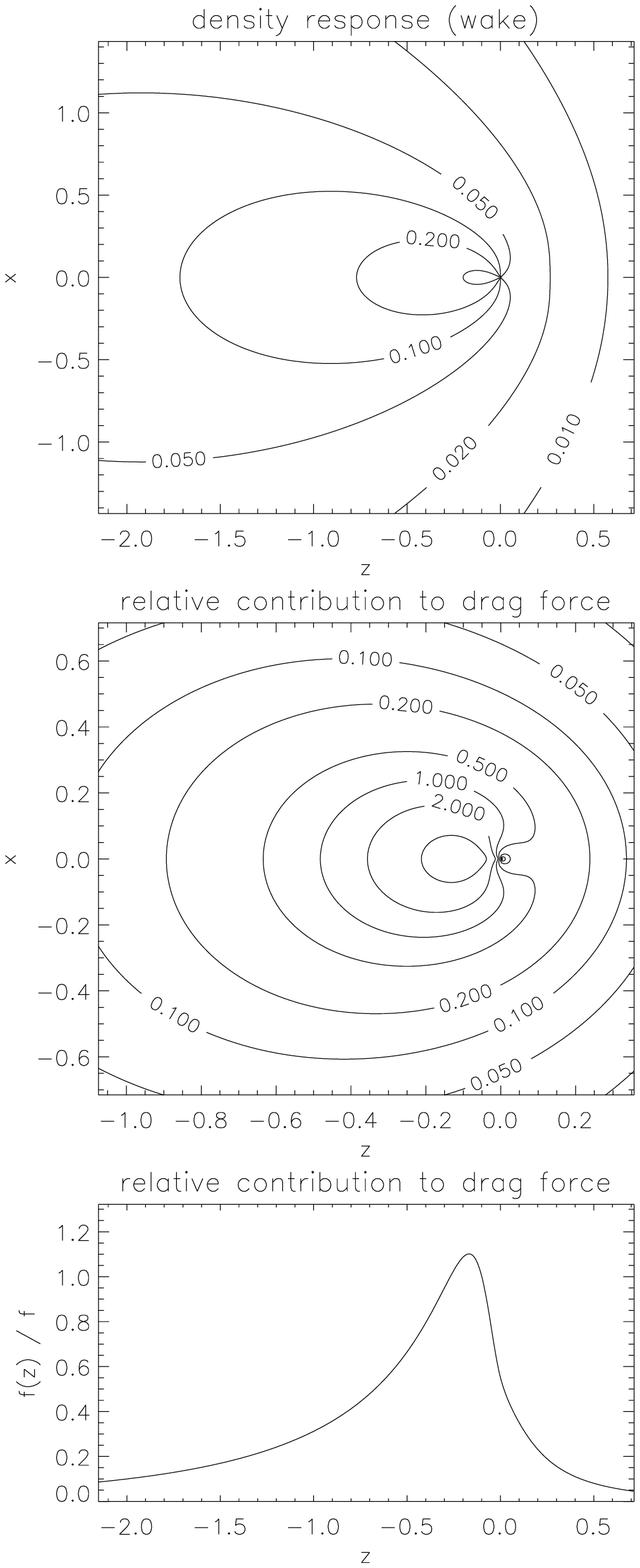,width=9.0cm,silent=1} \vskip -2.0cm}
\caption{{\bf Figure 1b.} Same as Fig. 1a, but for an extended subhalo,
with $R_{\rm h}/R_{\rm s}=60$. The wake is clearly more extended than
for the compact subhalo, with a peak which has a size on the order of
the size of the subhalo. This is also true for the compact subhalo
(See Fig. 1a).}
\endfigure

\subsection{Structure of the wake}

An important issue addressed in this paper is how well dynamical friction can
be modelled using N-body simulation techniques. In order to assess this,
we need to establish the volume and shape of the {\it effective}
gravitational wake that generates the drag force, and thus the actual
number of N-body particles that are available to model this wake.
For this purpose we employ to formalism of Danby \& Bray (1967),
which allows one to calculate the density response of any extended
subhalo moving through a uniform background density field.

The difference between compact and very extended subhaloes is shown in
Figs. 1a and 1b, in which the density response $\rho_{\rm res}(x,y,z)$ is
plotted in the $x-z$ plane, with the subhalo moving along the $z-$axis
(top panels).
In order to establish the relative importance of each spatial position
towards the cumulative dynamical friction force, first note the cylindrical
symmetry of the problem. This means that we only need to consider
the drag force along a circle at $(R,z)$, with $R^2=x^2+y^2$ being the
cylindrical coordinate defined as the distance to the $z$-axis.
The contribution to the total drag force on an extended spherical
subhalo at $(0,0)$ from the density response at $(R,z)$ is
$\rho_{\rm res}(R,z) M_{\rm s}(<r) / r^2$, with $r^2=R^2+z^2$.
For Plummer subhaloes this is equal to
$\rho_{\rm res}(R,z) M_{\rm s} r / (R_{\rm s}^2 + r^2)^{3/2}$.
For the two specific subhaloes considered this quantity is shown in
the corresponding middle panels of Figs. 1a and 1b, again in the $(x,z)$-plane,
but zoomed in by a factor of two with respect to the top panels.
Integrating over $R$, we obtain the relative contribution to the
drag force as a function of $z$ only, which is shown in the lower
panels of Figs. 1a and 1b.

What Figs. 1a and 1b demonstrate is that the contribution to the drag force
acting on a compact subhalo comes mostly from a compact region just behind
the subhalo, whereas for more extended subhaloes this region is
more extended as well. It is clearly much harder to form a compact wake
within an N-body simulation with limited resolution at the subhalo scale.
The figures also demonstrates a point already made by Mulder (1983), which is
that the density response contours are identical well away from the subhalo,
with the only difference being their amplitude. Thus, if the narrow peak of the
wake behind the compact satellite is not formed due to a lack of particles,
the density response of both the compact and extended subhaloes will look
identical with a much smaller difference in amplitude than predicted by
Chandrasekhar's formalism.


\beginfigure*{3}
{\vskip 0.4cm \psfig{file=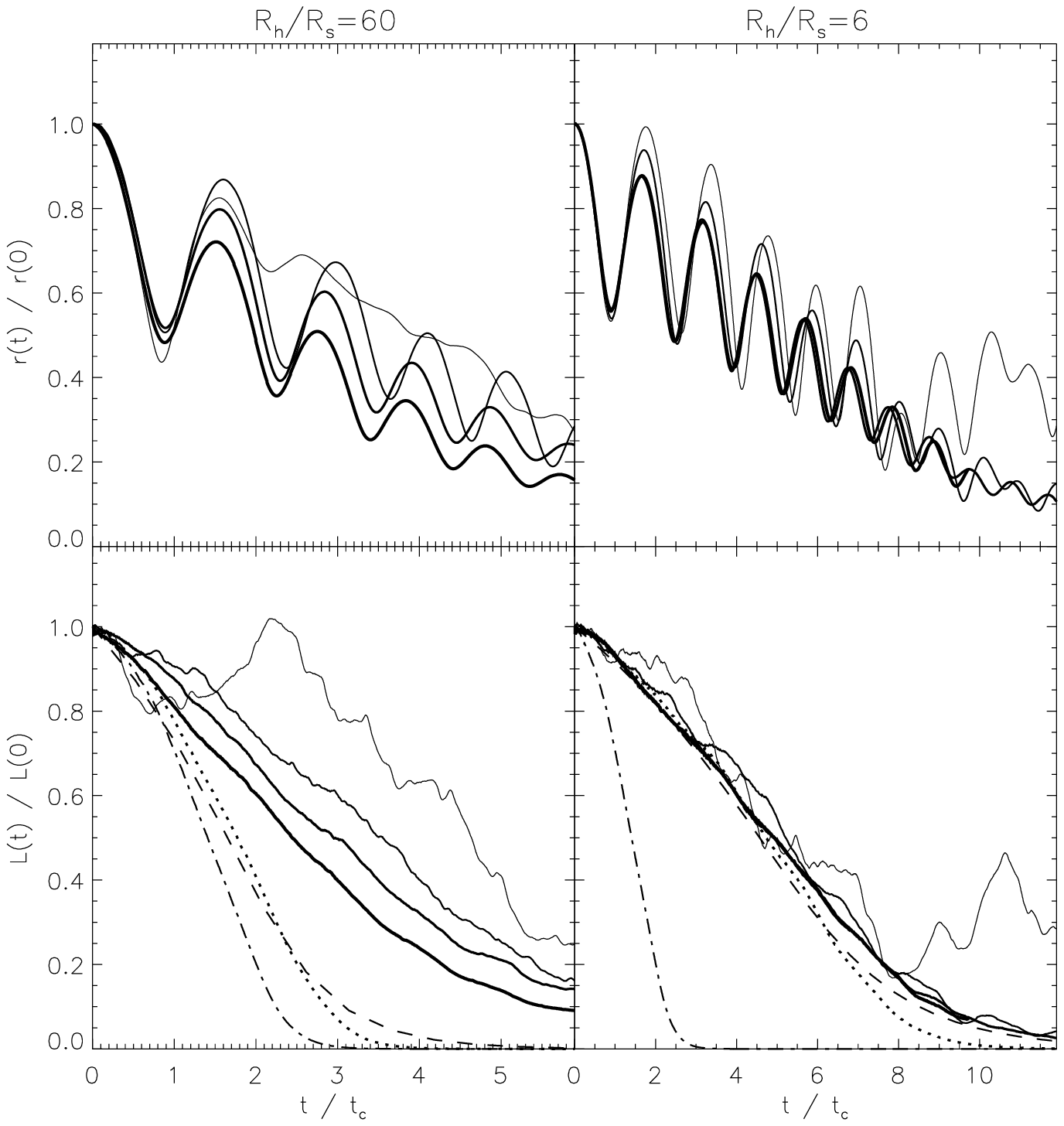,width=16.5cm,silent=1} \vskip -0.8cm}
\caption{{\bf Figure 2.} Time evolution of the radius and anugular
momentum of the slightly eccentric orbit ($\eta=0.76$, see main text) of a
subhalo initially at the half-mass radius of its embedding halo. In each
panel the four solid lines represent the N-body simulations, with the
thickness of the line increasing with the number of particles, being $10^3$,
$10^4$, $10^5$, and $10^6$ particles (i.e. the thickest solid line represents
the highest resolution simulation). The dotted and dashed line are
theoretical predictions, calculated in two different ways (see text), while
the dot-dashed line gives the theoretical prediction for a point mass,
representing the fastest decay by dynamical friction that is possible.}
\endfigure


\beginfigure{4}
{\psfig{file=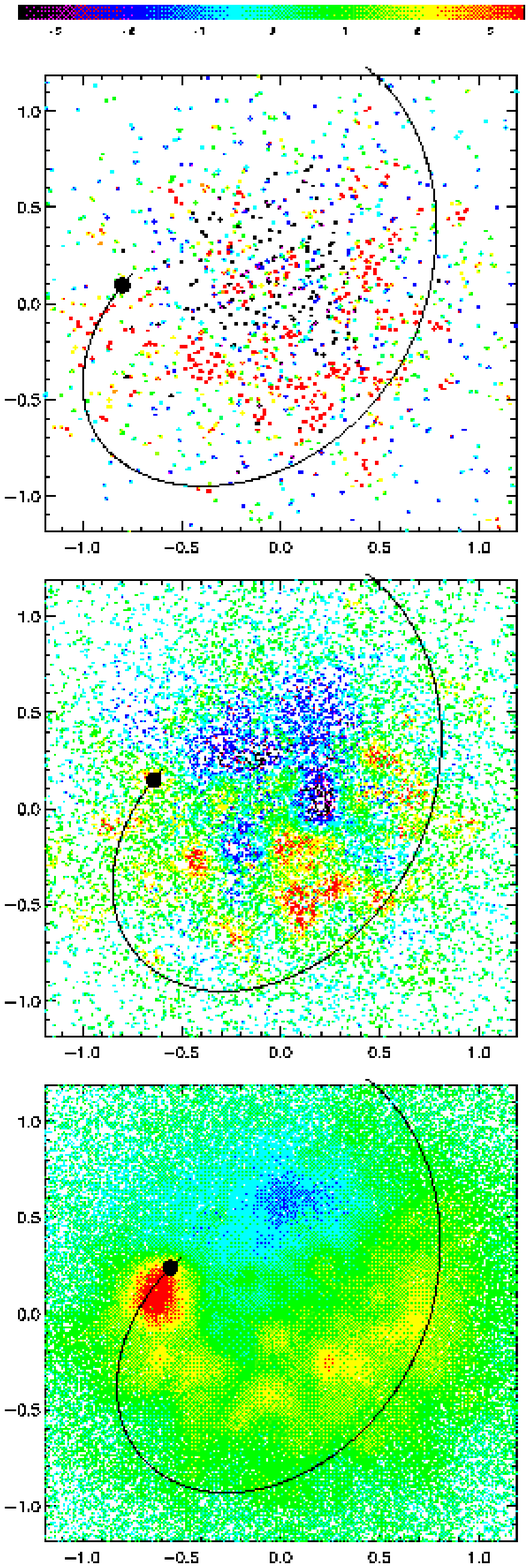,width=9.35cm,silent=1}}
\caption{{\bf Figure 3a.} The density response due to a compact subhalo
for three different numerical resolutions: $10^4$ (top panel),
$10^5$ (middle panel), and $10^6$ (bottom panel) particles for the
parent halo.}
\endfigure


\beginfigure{5}
{\psfig{file=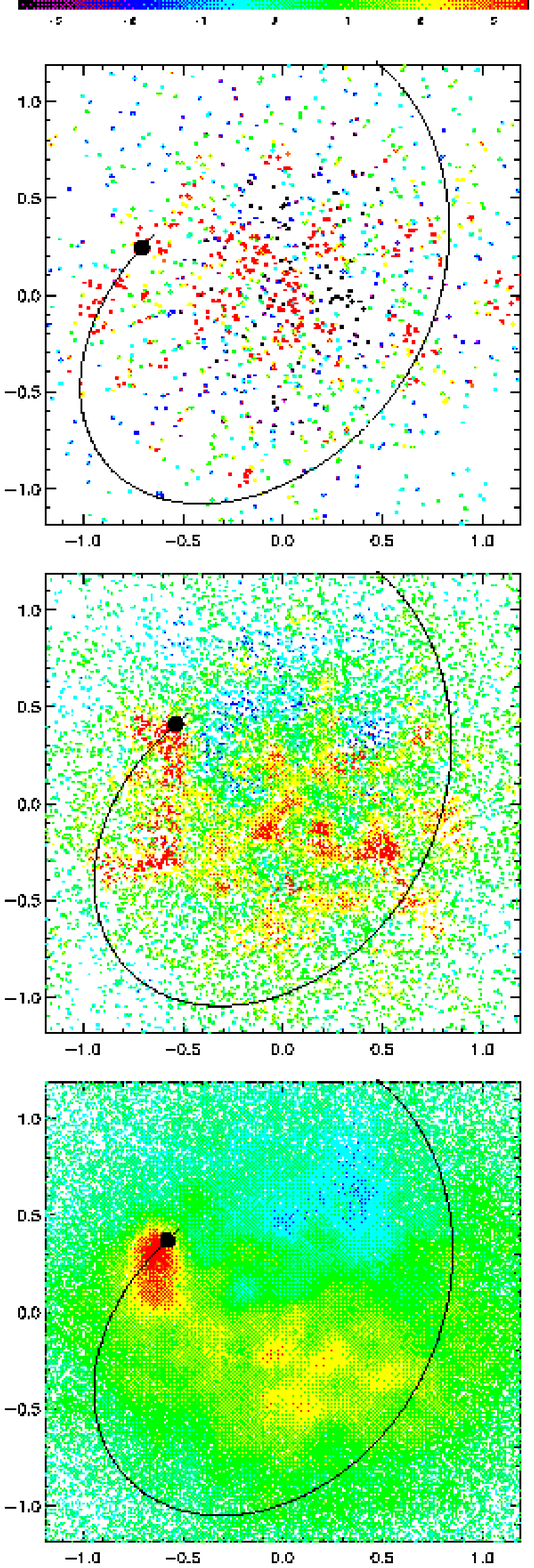,width=9.35cm,silent=1}}
\caption{{\bf Figure 3b.} As Fig. 3a, but for an extended subhalo.}
\vskip 0.1cm
\endfigure

\beginfigure*{6}
{\vskip 0.4cm \psfig{file=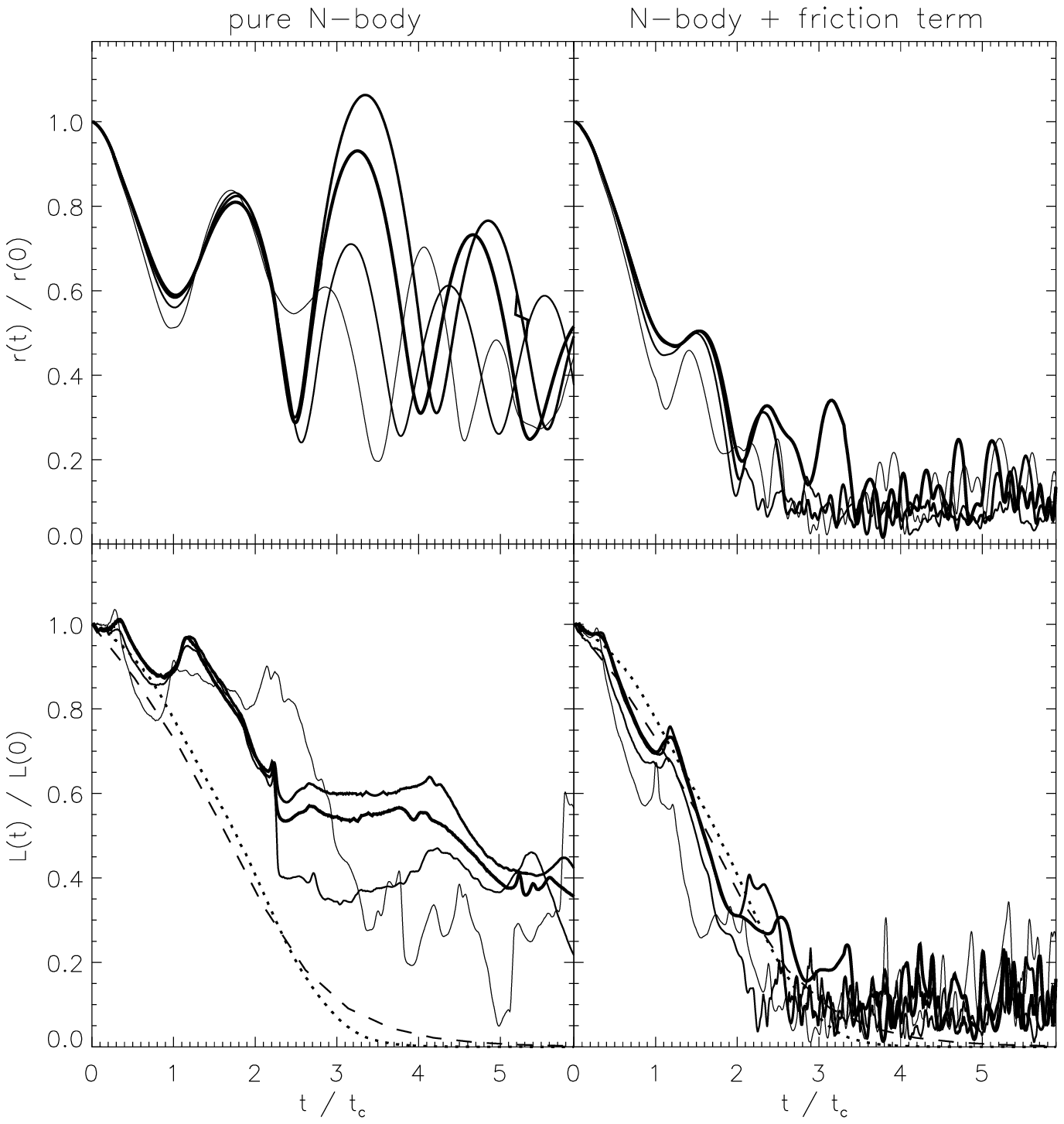,width=16.5cm,silent=1} \vskip -0.8cm}
\caption{{\bf Figure 4.} Same as Fig. 2, for the $R_{\rm h}/R_{\rm s}=60$
subhalo, which now shares its parent halo with 20 more subhaloes of the same
type. The left-hand panels show the result of re-running the same
simulation that produced the results shown in the left-hand panels of Fig. 2.
The right-hand panels shown the result of explicitely adding to the
equations of motion the local dynamical friction force according to
Chandrasekhar (1943).}
\endfigure

\beginfigure*{7}
{\vskip 0.4cm \psfig{file=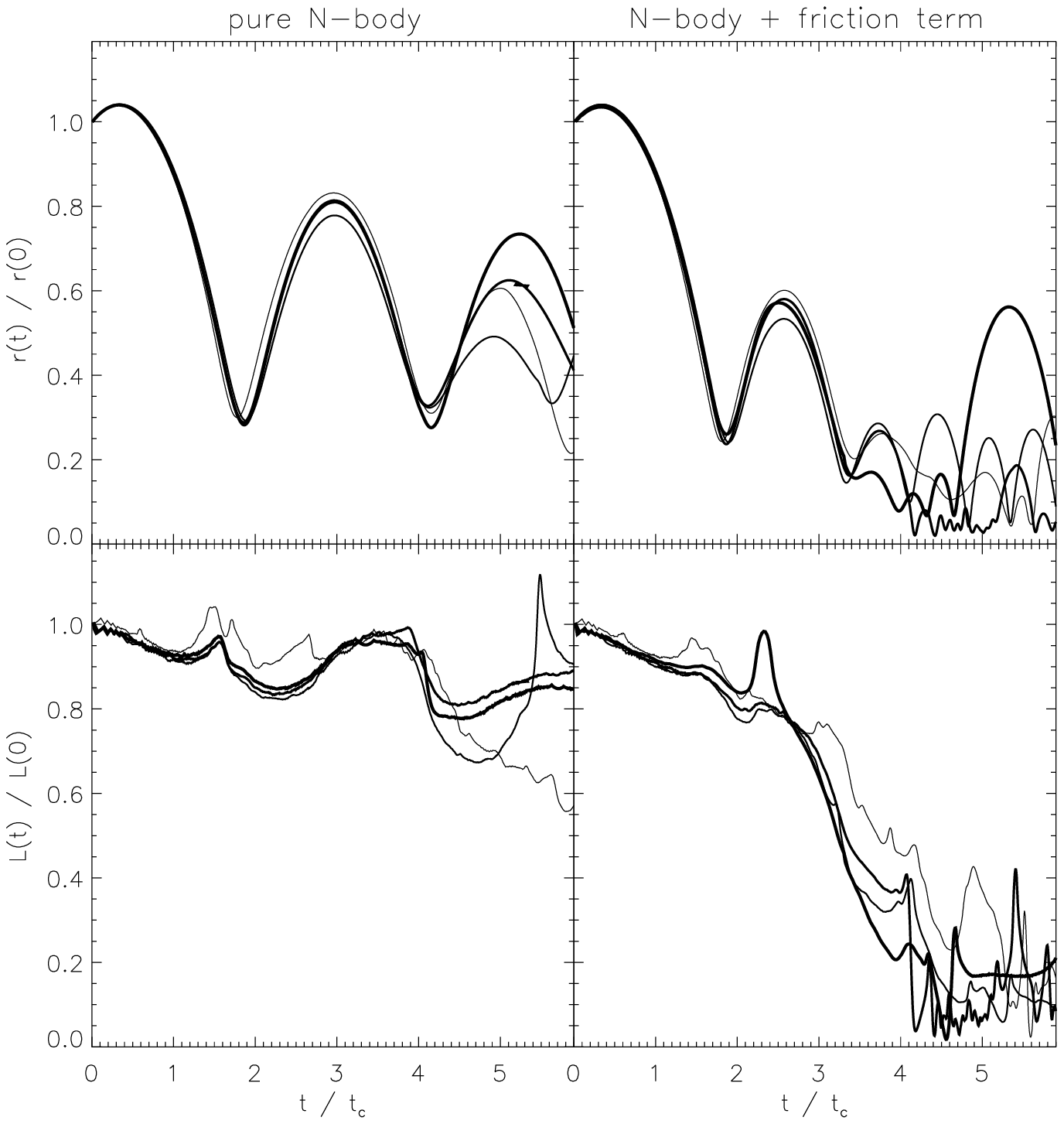,width=16.5cm,silent=1} \vskip -0.8cm}
\caption{{\bf Figure 5.} Same as Fig. 4, but for another subhalo (one of the
twenty that were added to the configuration used to produce Fig. 2).}
\endfigure

\subsection{Single subhalo simulations}

In the first series of test models we introduce at the half mass radius
a single, rigid subhalo with a mass of 1 per cent of the mass of the
embedding halo. Its orbit is mildly eccentric, with orbital circularity
$\eta=0.73$ (see Section 2.3.3 for the definition of $\eta$). 
This series represent the `classical' dynamical friction
test, i.e. a single subhalo decaying towards the centre of a halo.
Both halo and subhalo are modelled by a Plummer density profile,
with characteristic radii $R_{\rm h}$ and $R_{\rm s}$ respectively.
For this configuration the Coulomb logarithm is given by
$$\ln\Lambda = {1\over 2}\Bigl[\ln(1+R^2_{\rm h}/R^2_{\rm s}) -
  {R^2_{\rm h}/R^2_{\rm s}\over 1 + R^2_{\rm h}/R^2_{\rm s}}\Bigr].
  \eqno\stepeq$$
(Cora et al. 1997). For $(R_{\rm h}/R_{\rm s})>4$ this relation
is well approximated by $\ln\Lambda=\ln(R_{\rm h}/R_{\rm s})-0.5$.
We set $R_{\rm h}/R_{\rm s}=60$, which implies $\ln\Lambda=3.6$.
This would lead to a dynamical friction time of 12 embedding halo
crossing times for an isothermal sphere (eq. \the\dynfric), but
for a Plummer model the subhalo decays in 3 to 5 embedding halo
crossing times. The main reason for this faster decay is that the
subhalo starts at the half-mass radius $r_{\rm h}$, which is close to
$R_{\rm h}$ for the Plummer profile: $r_{\rm h}\approx 1.3 R_{\rm h}$.
Thus, it is already near the core of almost constant density, which
never happens for the core-less isothermal profile. The Plummer density
profile was chosen for the test runs because the large core assures that
the system behaves well numerically. Also, the particle density will
be larger for a larger volume of the simulation than for an isothermal
profile, so that the number of particles inside the wake will also be
larger. This does mean that if it is not possible to properly simulate
dynamical friction for a Plummer model, it is certainly not possible for
an isothermal or other core-less halo.

The dynamical friction time for an isothermal halo is well-defined,
being by the angular momentum evolution $L(t)\sim -t^{1/2}$
(e.g. Binney \& Tremaine 1987, from their eq. 7-25).
However, this is not the case for the Plummer halo,
for which $L(t)$ evolves fast near $R_{\rm h}$, but slower near its centre.
An analytical expression for $L(t)$ for a Plummer halo does not exist,
so $L(t)$ was solved numerically, using two quite different methods.
For the first method the calculation of Binney \& Tremaine (1987) for
the isothermal halo is adapted for a Plummer halo, but the more complicated
equation of motion is solved numerically. The second method is even more
numerical, as a special N-body code is used to solve the two-body
halo-subhalo system, in which the dynamical friction deceleration given
by eq. (\the\dynfricterm) is explicitly added to the equations of motion
coded in the N-body simulation.

We show $r(t)$ and $L(t)$ of the subhalo as a function of numerical
resolution in the left-hand panels of Fig. 2 (solid lines), where
increasing thickness of the lines indicate increasing resolution.
The two theoretical solutions for $L(t)$ for the subhalo are shown as
a dashed line (numerical solution to the Binney \& Tremaine formalism)
and a dotted line (the special N-body code). The dot-dashed line shows
the theoretical evolution of $L(t)$ for a point mass subhalo,
which represent the fastest decay possible for a subhalo of this mass
and initial orbit, i.e. $\ln\Lambda=\ln(m_{\rm h}/m_{\rm s})=4.6$.
At all resolutions we see the orbit of the subhalo decay, but this
happens increasingly faster for simulations with increasing resolution.
More important, however, is the observation that the $L(t)$ curves
converge towards the theoretical curve, but do {\it not} match it even
for $10^6$ particles. Thus, the subhalo needs
twice as long to decay even in the highest resolution simulation.
The most obvious explanation is that the gravitational wake that drives
dynamical friction is not properly modelled due to insufficient resolution.

In order to visualize this, we look at the density response of the embedding
halo in three of the runs from the first series (for $10^4$, $10^5$, and
$10^6$ particles).
First, the small-scale density was calculated using a Gaussian filter of
30 kpc, where the subhalo particle was excluded.
Then, the Plummer density law was fitted to the particles of embedding halo,
again excluding the subhalo. The resulting {\it response density}
i.e. the difference of the simulated density and the smooth Plummer law,
is shown in Fig. 3a.
The particles in a thin slice within the plane of the subhalo's
orbit are plotted, with a colour coded level indicating the response
density. The position and velocity of the subhalo after two embedding
halo crossing times, are indicated by a large black dot and a thick
line-segment respectively. Its orbit is shown as a thin solid line.
The wake can clearly been seen in the bottom panels, but is
absent in the top panels.

However, before pointing at insufficient resolution as the cause for
the slower than expected decay of subhaloes in N-body haloes, we should
look at a configuration where resolution should be less of a concern. For this
purpose we ran a second series of simulations, completely identical to the
first one, in which the subhalo is taken to be 10 times larger in extent,
but with the same mass and initial orbit. Setting $R_{\rm h}/R_{\rm s}=6$,
we get $\ln\Lambda=1.3$, which is also used to calculate (numerically) two new
theoretical predictions for $L(t)$.
The results are plotted in the right-hand panels of Fig. 2,
for the same range in particle number as the first run.

At all resolutions, we see the subhalo spiralling towards the centre
at roughly the same rate, except for the $10^3$ particle run, which clearly
shows two-body interactions to be important all the way. The decay is
slower than for the more compact subhalo modelled in the first run, as
it should be, but most interestingly, it is a fair match to the theoretical
curve. Furthermore, a wake, plotted in Fig. 3b,
is visible not only for the $10^6$ particle run, but also for the $10^5$
particle simulation, and maybe even for the $10^4$ particle run.

Thus, for a subhalo this size, dynamical friction seems to be modelled
properly using as little as $10^4$ particles for the embedding halo.
However, it is still not certain whether the gravitational wake is actually
modelled properly, because global dynamical friction could also drive
part or even most of the decay, and the wakes visible in Figs. 3a and 3b might
well be weaker than predicted. If this is the case, the match with
Chandrasekhar's prediction is coincidental.

Besides the gravitational wake, Figs. 3a and 3b also show the global
distortion induced by the subhalo. What is important is that the
distortions can be seen for {\it all} resolution. This means that
global dynamical friction is clearly an active process in an N-body
simulation of a halo with a single subhalo, irrespective of numerical
resolution. It is quite possible that global dynamical friction drives
most of the decay of the orbit of this single subhalo. It is thus essential
to see what happens for multiple subhaloes.


\subsection{Multiple subhaloes simulations}

A shortcoming of the `classical' dynamical friction test is that just
a single subhalo is considered, which is not realistic for hierarchical
structure formation scenarios. It is also not a proper test for
Chandrasekhar's dynamical friction formula, as global tidal distortions
(as visible in Figs. 3a and 3b) drive orbital decay as well, which is not
likely if many subhaloes orbit the same halo (as discussed in Section 4.1).
This leads us to perform a third series of simulations, in which
the first subhalo is taken to be identical to the subhalo of the first
series, but twenty more subhaloes are added, on random orbits.

Each of the subhaloes can only decay through local dynamical friction,
with a timescale given by the Chandrasekhar formalism (eq. \the\dynfric),
because global distortions from multiple subhaloes form a net stochastic
force which both accelerates and decelerates each subhalo, without the
net effect of decay. Even more stochasticity comes from the sum of forces
from all subhaloes and their wakes.
These forces should average out over time, so each subhaloes decays
only under the influence of their own wake, which will be the closest
overdensity at most times.

The evolution of radius and angular momemtum for the original subhalo
are plotted in the left-hand panels of Fig. 4. Their evolution clearly shows
the influence of the stochastic forces discussed above. The orbit
decays only some of the time, and less rapidly. Also, the subhalo does
not reach the centre of it parent halo after six crossing times,
as it did in the single subhalo simulation (see Fig. 2). Some of the
added subhaloes do not decay at all, even though they start at roughly the
same radius. This is shown in the left-hand panels of Fig. 5, which again
shows the evolution of radius and angular momentum, but for one of the
added subhaloes.

If the density response were properly modelled, local dynamical friction
would bring the subhaloes towards the centre of their parent halo within
a few crossing times. In order to show this, the friction force is explicitely
added to the equations of motion in the N-body code. The local density
is calculated using Guassian smoothing of the particle distribution,
and the velocity distribution is assumed to be Maxwellian, which is true
for the test models considered here, but not for general cosmological mass
distributions. The results of re-running the simulations with the local
friction term added are shown in the right-hand panels of Figs. 4 and 5, for
the same two subhaloes shown in the left-hand panels of these figures.

Not surprisingly, the two subhaloes, and indeed all other subhaloes,
decay at the predicted rate. The point of this exercise is to show that
if dynamical friction were properly modelled, it is very effective in
destroying subhaloes. However, unless a drag force is explicitly
added, even a halo of $10^6$ particles is not capable of modelling
local dynamical friction self-consistently. Dynamical friction is only
effective in numerical simualations of a single subhalo system, where
global dynamical friction drives the orbital decay.

\subsection{What is needed to properly model dynamical friction ?}

How many particles are required to properly model the gravitational
wake that drives (local) dynamical friction ? From the test above it seems
that the answer is at least $10^8$ particles for single subhalo systems,
but probably a lot more, as the multiple subhalo simulations
demonstrate. The main problem is the lack of particles near the subhalo
to form a strong and sustainable wake, especially away from the
centre of the embedding halo where the N-body particle density is low.
The extent of the wake is determined by the extent of the subhalo
and its velocity with respect to the rest-frame of its parent halo,
in the sense that the wake is more extended with a lower maximum
responds density for a more extended and/or slower subhalo (Mulder
1983). Thus, for the more compact subhalo of the
first series, the wake is more linear, aligned along the orbit
of the subhalo. This wake consists of fewer particles, which responded
stronger to the compact subhalo than to the extended one, and 
the compact subhalo wake is therefore more fragile numerically.
The problem is worse in the outskirts of the embedding halo,
where the N-body particle density is lowest and the
response density peak therefore more sparsely sampled,
and for large subhalo velocities, especially for a subhalo that
just entered a halo. In the latter case the wake is quite narrow
(see Fig. 2b of Mulder 1983), and again hard to sample using the
available particles from the embedding halo.

Another numerical problem is that two-body interactions can
disrupt a wake, just like small subhaloes are disrupted by two-body
interactions (see Section 2.2). Can we use eq. (\the\twobodyheating)
to estimate a wake disruption timescale ? This seems unlikely, as the
wake is not a simple collapsed structure.
An N-body particle that is part of a wake only remains so for some time.
During this time, the only two-body interaction
such a particle should have is with the subhalo, and not with other
particles in the wake. Thus, the number of particles in and
around the wake should be large enough to prevent this. The simulations
as performed show this: for the more extended subhalo in the second
series of test simulations the volume of the wake is fairly wide,
and thus the number of particles is sufficient even for the
$N_{\rm h}=10^4$ run. But for the compact halo, the wake has a compact
density response maximum, which seems easy to disrupt by two-body effects.

Thus, if far too many particles are required to simulation dynamical
friction self-consistently, the solution is to explicitly add the
drag force to the equations of motion, as was done for the fourth series
of test simulations presented above. However, this was easy to implement
for the Plummer halo, as its properties could be coded directly and
the velocity distribution function is known, but this will be much harder
for cosmological N-body simulations in which haloes can have any
profile and typically are not smooth and relaxed, and subhaloes are not
single particles. It will therefore require some ingenuity to explicitly
include the dynamical friction force in a cosmological N-body code.


\section{Timing}

In hierarchical clustering, haloes grow through merging with other
haloes. This growth is not continuous, but occurs in distinguishable
merger events, with relatively quiet periods in between. During these
periods substructure, including subhaloes, can be destroyed by
the physical processes described in Section 2.3, and in simulations
also by the numerical processes described in Section 2.2.
When comparing the survival of subhaloes in galaxies and in clusters,
the epoch and duration of these quiet intervals between merger events
surely must play an important r\^ole. In order to investigate this
we first estimate the mean formation epoch of a halo of a given
mass $M$ (or corresponding circular velocity $v_{\rm c}$).

\beginfigure{8}
{\psfig{file=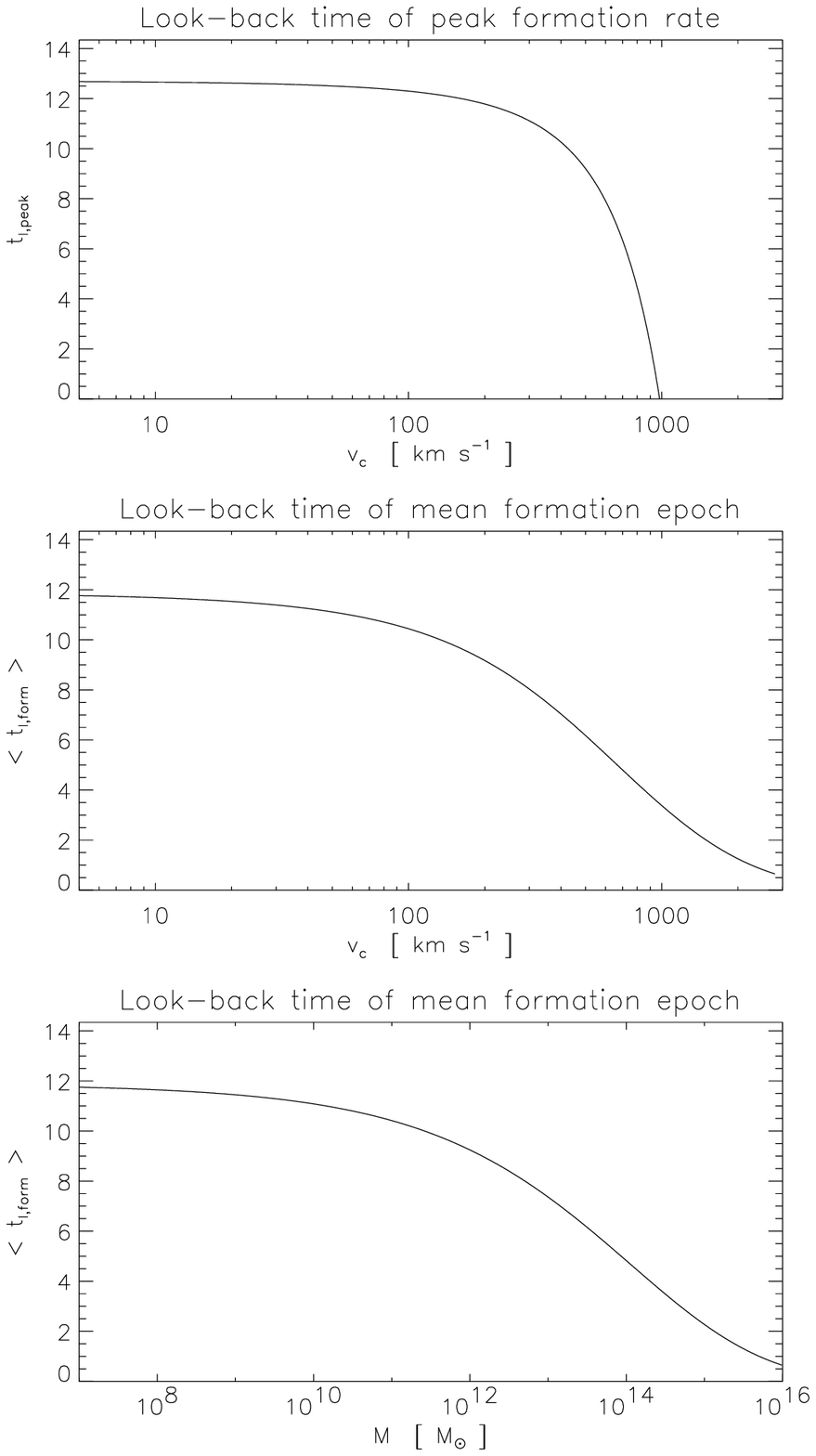,width=8.5cm,silent=1}}
\caption{{\bf Figure 6.} The top panel shows the look-back time of the {\it peak}
formation rate of a halo with a given circular velocity.
The middle and bottom panels display the same for the {\it mean} formation
epoch, as a function of halo circular velocity and halo mass respectively.}
\endfigure

\subsection{Halo formation epoch}

The Press-Schechter formalism gives a good description of the collapse of
haloes, and therefore of their formation rate. For the special case
of an Einstein-de Sitter universe the halo formation rate is given by
$${{\rm d} n(M)\over {\rm d} t} \sim (1+z)^{7\over 2}
  e^{-\delta^2_{\rm c}(1+z)^2/(2\sigma^2(M))} \eqno\stepeq$$
(Percival \& Miller 1999), where $n(M)$ is the number density of haloes,
$\delta_{\rm c}\approx 1.67$ is the critical overdensity for collapse, 
and $\sigma(M)$ is the variance of the density field filtered with a
sharp $k$-space filter corresponding to the mass-scale $M$.
The peak formation epoch is given by
$$z_{\rm peak}(M) =
   \Bigl({7\sigma^2(M) \over 2\delta^2_{\rm c}}-1\Bigr)^{1\over 2}
  . \eqno\stepeq$$
The spherical collapse model can be used to assign a circular
velocity to a halo of mass $M$ forming at redshift $z$:
$$v_{\rm c} = \Bigl({M\over 2.35\times 10^5 h^{-1}
        {\rm M}_\odot}\Bigr)^{1\over 3}
	(1+z)^{1\over 2}\ {\rm km s}^{-1} \eqno\stepeq$$
(e.g. White 1996).
Instead of $z_{\rm peak}(M)$, we plot the corresponding look-back time
$t_{\rm l,peak}$ (with $t_{\rm l}=t_0-t$, and $t_0$ the age of the Universe)
as a function of $v_{\rm c}$ in the top panel of Fig. 6,
for the standard CDM spectrum normalized to $\sigma_8=0.67$.

The {\it mean} formation redshift is obtained by integrating over
time up to but not beyond the present epoch $t_0$:
$$\langle z_{\rm form}(M) \rangle = \Bigl[\int_0^{t_0} n(M) t dt\Bigr] / 
   \Bigl[\int_0^{t_0} n(M) dt\Bigr]\ . \eqno\stepeq$$
Again, the corresponding look-back time $t_{\rm l,form}(v_{\rm c})$
is plotted in the middle panel of Fig. 6. For convenience,
$t_{\rm l,form}(M)$ is shown in the bottom panel of Fig. 6.

\subsection{Available time for subhalo disruption}

Having established when haloes form through merging of
small haloes, one can then estimate how much time is available
to the new halo to destroy its subhaloes.
From the mean formation look-back times shown in Fig. 6
it is straightforward to see that dwarf galaxies with
$v_{\rm c}=10-20$ km s$^{-1}$ form 11.5 Gyr ago, galaxies with 
$v_{\rm c}=100-200$ km s$^{-1}$ form $9-10.5$ Gyr ago, whereas
galaxy clusters with $v_{\rm c}=1000-2000$ km s$^{-1}$ form on
average 1-3 Gyr ago. This means that dwarfs falling into galaxies
had, on average, about {\it four times} more time to be disrupted
than galaxies falling into clusters.

\subsection{Subhalo replacement rates}

Even if subhaloes are efficiently destroyed by their embedding halo,
be it a galaxy or a galaxy cluster, new subhaloes can fall
in and replace the destroyed ones. This is certainly happening
for galaxy clusters, for which the peak formation epoch is
right about now (see top panel of Fig. 6). However, roughly
half the galaxies formed about 9 Gyr ago, and very few are
forming at the present epoch. Thus, continued growth by
secondary infall, as seen for galaxy clusters, is almost
absent. Thus, even if the destruction rate for subhaloes
in galaxies and clusters are exactly the same, cluster galaxies
are being replaced in fair numbers, while galaxy subhaloes are not.

\subsection{Dynamical friction during hierarchical structure formation}

An important factor in the efficiency of dynamical friction
is the mass ratio of the embedding halo to its subhalo,
$m_{\rm h}/m_{\rm s}$. Taken at face value, this means that
a $10^9$ M$_{\sun}$ dwarf within a $10^{12}$ M$_{\sun}$
isothermal galaxy decays in of order 40 Gyr (see Section 2.3.3).
However, such a dwarf is likely first spend some time
in a smaller galaxy, before ending up in the $10^{12}$ M$_{\sun}$
galaxy. For example, assume that through hierarchical merging
it spends 1 Gyr in a $10^{10}$ M$_{\sun}$ galaxy, then 2 Gyr
in a $10^{11}$ M$_{\sun}$ galaxy, and finally 6 Gyr in a
$10^{12}$ M$_{\sun}$ galaxy. The dynamical friction times
are then, respectively 0.4, 4, and 40 Gyr. Thus, this example
subhalo will already be destroyed by the $10^{10}$ M$_{\sun}$
galaxy before the latter grows to a $10^{11}$ M$_{\sun}$ galaxy.
Here we have not taken into account that the crossing time
of the $10^{10}$ M$_{\sun}$ galaxy at look-back time
$t_{\rm l}=9$ Gyr is typically smaller than for the final
$10^{12}$ M$_{\sun}$ galaxy, and that its mean density is larger
than for a $10^{10}$ M$_{\sun}$ galaxy that formed recently,
which will speed up dynamical friction.

This example illustrates that the very nature of
hierarchical clustering within an expanding Universe implies that
dynamical friction is much more efficient when taking into account
the growth of the embedding halo and the expansion of the Universe
(i.e. densities are higher at higher redshifts).

\section{Summary, discussion and conclusions}

This paper considered the various numerical and physical processes
driving the disruption of subhaloes, with the specific aim to find out
whether subhaloes can survive within galaxies, as they do in galaxy
clusters. This was prompted by the results of Klypin et al. (1999b) and
Moore et al. (1999), who found a large discrepancy between simulations
and observations with regard to the abundance of galaxy subhaloes. 

Two-body heating is the most efficient of the numerical processes,
but only effects small subhaloes, i.e. of order 100 particles or less.
Thus, given enough particles, the overmerging problem is resolved, and
only physical processes can operate on subhaloes. But do they ?
In Section 3 it was shown that subhalo-subhalo heating is not an
important process (contrary to the claim of Moore et al. 1996),
while the mean tidal field of an
embedding halo only tidally limits subhaloes, but generally does not
destroy them completely. A surprising finding of this paper is that the
physical process capable of completely destroying subhaloes, dynamical
friction, is not properly modelled by the N-body simulation technique.
It was found that in order to properly model the gravitational wake that
produces the dynamical friction drag force, one needs a much higher
resolution than can be achieved at present.

For single subhalo systems, the problem does not manifest itself because
part of the orbital decay is due to global distortions induced in the
embedding halo by the orbiting subhalo. This `global' dynamical friction
adds to the `local' dynamical friction produced by the gravitational wake.
Test simulations in which the resolution was increased from $10^3$ to
$10^6$ particles (for the embedding halo) showed that the subhalo angular
momentum decreased increasingly faster with particle number, but only in
a logarithmic fashion, and without any sign of convergence to the expected
rate. The problem does show for multiple subhalo systems, in which global
dynamical friction does not operate. Thus, only local dynamical friction
acts on each of the subhaloes, but even at the highest resolution presently
feasible, the wake is not nearly as strong as it should be, and orbital
decay is far too slow. Still, increasing the number of simulation particles
results in faster decay, so there exists a (very high) number of particles
for which the gravitational wake is properly simulated. As this is not the
case for current simulations, too few subhaloes are destroyed, and due to
the insufficient resolution there is in effect an undermerging problem.

All this means that there are five possible causes for the galaxy subhalo
problem:

(1) the initial density fluctuation spectrum cuts off near the dwarf
    galaxy scale

(2) dwarf galaxies are much darker than galaxies due to strong feedback

(3) overmerging produces too many dwarf galaxy haloes (except within more
    massive haloes, of course) in simulations with particle masses of order
    $10^6$ M$_{\sun}$

(4) dynamical friction is not properly simulated even at the highest
    numerical resolution achieved to date (undermerging)

(5) due to the timing of hierarchical formation and merging of haloes,
    galaxy subhaloes are more easily destroyed than cluster subhaloes

In order to solve the galaxy subhalo problem, at least one or more needs
to be true. The simplest but most radical solution is for cause (1) to be
true in the extreme, i.e. there is no power on small scales so that 
very few dwarfs galaxies actually form during a Hubble time.
Another simple solution is to assume that a significant feedback mechanism
operates in dwarf galaxies, so that mass-to-light variations alone can
explain the deficiency in observed dwarfs. But then
there remains a large abundance of dark dwarf galaxy {\it subhaloes}, which
pose a problem to the survival of a stellar disk. Clearly, it is
perfectly acceptable to have plenty of dark dwarfs in the field, but only
if they are efficiently destroyed after becoming part of a (bigger) galaxy
halo.

This plausible picture is not supported by the simulations of
Moore et al. (1999) and Klypin et al. (1999b). However, the results of
the test simuations presented in this paper suggest that
numerical shortcomings provide the answer, with the main culprits
being undermerging, which is the inability of N-body simulations
to properly simulate the gravitational wake that drives dynamical friction,
and overmerging of the smallest haloes, due to two-body interactions.
The consequence of undermerging is that subhaloes are not detroyed in
sufficient numbers, while overmerging overproduces them. This produces an
artificially high abundance of subhaloes only if overmerging operates on a
smaller mass-scale than undermerging, which is the case for simulations with
of order $10^6-10^8$ particles. This includes the simulations of
Moore et al. (1999) and Klypin et al. (1999b).

The numerical problems are especially bothersome
during the early stages of hierarchical structure formation,
where both haloes and subhaloes are modelled by relatively few particles.
Galaxy haloes are, on average, four times older than galaxy
cluster haloes, and without much ongoing secondary infall. Subhaloes that are
destroyed are not all replaced; in this picture, the Magellanic Clouds are
relatively new to our Galaxy, and will be destroyed within a few Gyr
(Tremaine 1976).

In concluding, it is likely that several of the five causes mentioned
above conjure to completely solve the apparent galaxy subhalo problem.
The main problem is that
dynamical friction is not properly simulated yet, even in the highest
resolution simulations to date, due to numerical limitations which,
paradoxically, result in an 'undermerging' problem. This is made worse by
the timing of halo formation and merging in hierarchical structure formation,
which favours destruction of subhaloes in galaxies over subhalo destruction in galaxy
clusters. Finally, recent galaxy formation models predict a larger
mass-to-light ratio for galaxy subhaloes than for galaxy cluster subhaloes.
The net effect of these three causes is that galaxies have a relatively
low abundance of subhaloes, i.e. dwarfs, while at the same time a large
number of field dwarf galaxies can exist which are dark enough to be missed
observationally.

\section*{ACKNOWLEDGEMENTS}

Many thanks to Frank van den Bosch, Ben Moore, John Peacock, and
Will Percival for fruitful discussions and comments.


%
%

\end